\documentclass[conference]{IEEEtran}
\usepackage{def}
\usepackage{package}
\usepackage{hyperref}

\usepackage{titlesec}
\titlespacing*{\section}
{0pt}{0.9ex plus 1ex minus .2ex}{0.9ex plus .2ex}
\titlespacing*{\subsection}
{0pt}{0.8ex plus 1ex minus .2ex}{0.8ex plus .2ex}

\def\BibTeX{{\rm B\kern-.05em{\sc i\kern-.025em b}\kern-.08em
    T\kern-.1667em\lower.7ex\hbox{E}\kern-.125emX}}

\title{\arch: A Network and Architecture Co-design for Offloading $\mu s$-scale Datacenter Applications}

\author{
\IEEEauthorblockN{Yifan Yuan$^{1}$, Jinghan Huang$^{1}$,Yan Sun$^{1}$, Tianchen Wang$^{1}$, Jacob Nelson$^{2}$, Dan R. K. Ports$^{2}$, \\
Yipeng Wang$^{3}$, Ren Wang$^{3}$, Charlie Tai$^{3}$, Nam Sung Kim$^{1}$} 

\IEEEauthorblockA{\textit{$^{1}$UIUC}, \textit{$^{2}$Microsoft Research}, \textit{$^{3}$Intel Labs}}
}

\begin{document}
\maketitle

\thispagestyle{plain}
\pagestyle{plain}

% !TEX root = paper.tex
\begin{abstract}
\footnote{This work has been accepted by a conference. The authoritative version of this work ``RAMBDA: RDMA-driven Acceleration Framework for Memory-intensive  us-scale Datacenter Applications'' will appear in the Proceedings of the 48th IEEE/ACM International Symposium on High-Performance Computer Architecture (HPCA-29), 2023.}
Responding to the ``datacenter tax'' and ``killer microseconds'' problems for datacenter applications, 
diverse solutions including Smart NIC-based ones have been proposed. Nonetheless, they often suffer from high overhead of communications over network and/or PCIe links. 
To tackle the limitations of the current solutions, 
this paper proposes \arch, a holistic network and architecture co-design solution that leverages current RDMA and emerging cache-coherent off-chip interconnect technologies.
Specifically, \arch consists of four hardware and software components: 
(1) 
unified abstraction of inter- and intra-machine communications managed by one-sided RDMA write and cache-coherent memory write; 
(2) efficient notification of requests to accelerators assisted by cache coherence; 
(3) cache-coherent accelerator architecture directly processing requests received by NIC; and  
(4) adaptive device-to-host data transfer 
for modern server memory systems consisting of both DRAM and NVM exploiting state-of-the-art features in CPUs and PCIe.
We prototype \arch with a commercial system and evaluate three 
popular datacenter applications: in-memory key-value store, chain replication-based distributed transaction system, and deep learning recommendation model inference. The evaluation shows that \arch provides 30.1$\sim$69.1\% lower latency, up to 2.5$\times$ higher throughput, and $\sim3\times$ higher power efficiency than the current state-of-the-art solutions. 
\end{abstract}

% !TEX root = paper.tex
\section{Introduction}
\label{sec:intro}

%The fast pace of datacenter network evolution increasingly places an enormous
The datacenter network has evolved at a fast pace.
Currently, 100 Gbps Ethernet is widely adopted by datacenters, and 400 Gbps Ethernet is not far on the horizon~\cite{bf-3}.
At these rates, a server may be tasked with processing hundreds of millions of
packets per second. However, single-thread performance of CPUs has remained
comparatively stagnant, requiring the CPUs to spend more cores and their cycles %an increasing fraction of CPU cycles/cores to be spent on 
for network processing -- a major component of the
``datacenter tax''~\cite{Kanev:2015:PWC:2749469.2750392}. 
%For some applications, offloading computation to accelerators can provide faster or more energy-efficient execution, but 
%but as identified in the ``killer microsecond'' problem~\cite{10.1145/3015146,8675221}, the small, $\mu s$-scale tasks which are common in modern datacenter applications~\cite{10.1145/3373376.3378450} make it challenging to benefit from accelerators. 
Besides, as identified in the ``killer microsecond'' problem~\cite{10.1145/3015146,8675221}, it is not suitable to offload these applications to conventional accelerators such as GPU as they are not efficient to process many small, $\mu s$-scale tasks which are common in modern datacenter applications~\cite{10.1145/3373376.3378450}. 
This makes offloading/acceleration of such $\mu s$-scale datacenter applications  challenging.

\begin{table*}
  \centering
  \caption{Taxonomy of hardware-based $\mu s$-scale datacenter applications offloading/acceleration.}
\scriptsize
  \label{tab:comp}
  %\begin{tabular}{m{3cm}m{1.5cm}m{1.5cm}m{1.5cm}m{1.5cm}}
  \begin{tabular}{m{7.7cm}ccccc}
    \toprule
   \bf Optimization &   \bf Net. Overhead & \bf PCIe Overhead & \bf CPU Overhead & \bf Flexibility & \bf Perf. Stability \\
    \midrule
  Two-sided RDMA/kernel-bypass with multi-core~\cite{199315,225980,196243,179747,10.1145/2806887,10.1145/2619239.2626299} & Low & Low & High & High & Low \\ 
  One-sided/mixed RDMA~\cite{drtm,179767,222609,10.1145/2901318.2901349,180191,10.1145/3319647.3325827,cell} & High & High & Low & Low & Low\\ 
  (Smart)NIC offloading~\cite{10.1145/3373376.3378528,li2017kv,179402,10.1145/3230543.3230572,222623,10.1145/3422604.3425923,10.1145/3342195.3387519,rdma-turing,10.1145/3477132.3483565,10.1145/3477132.3483555,10.1145/3477132.3483587,smartnic3,smartnic1} & Low & High & Low & High & Low\\
  %Specialized integrated hardware~\cite{10.1145/3445814.3446696,10.1145/3352460.3358278,10.1145/3297858.3304070,10.1145/2749469.2750415,10.1145/2541940.2541965,273715,10.1145/2485922.2485926} & Low & Low & Medium & Low & High\\ 
  \bf\arch & \bf Low & \bf Low & \bf Low & \bf High & \bf High\\ 
    \bottomrule
  \end{tabular}
\end{table*}

Current solutions for the two problems %accelerating  $\mu s$-scale datacenter applications 
offload network and/or application processing from the server CPU using one of the three strategies listed in \tabref{tab:comp}.
The first strategy is kernel-bypass networking, including user-space stack~\cite{dpdk,arrakis,ix,mtcp} and two-sided Remote Direct Memory Access (RDMA)~\cite{199315,225980,196243,179747,10.1145/2806887,10.1145/2619239.2626299}. It reduces
the performance overhead imposed by packets going through the kernel space by delivering the data directly to the user space. Nonetheless, the server CPU still needs to do network processing from the user space (\ie, user-space stack) and application processing (\ie, both user-space stack and two-sided RDMA) that still
demand many CPU cores to achieve required performance~\cite{199315,225980,196243,179747,10.1145/2806887,10.1145/2619239.2626299}. 
The second strategy is one-sided RDMA. It allows clients to do application processing as they can bypass the server CPU and  
directly read from or write to the server memory.
%, an approach widely used for distributed systems~\cite{drtm,179767,222609,10.1145/2901318.2901349,180191,10.1145/3319647.3325827}.
%While in principle RDMA can completely bypass the server CPU, 
However, the limited semantics of one-sided RDMA operations 
require multiple network round trips to serve a single request from a client. For example, Pilaf~\cite{180191} performs a key-value lookup by reading a remote hash table to retrieve a pointer, then placing a second RDMA
request to read the corresponding data. 
The third strategy addresses
this problem by using a Smart NIC that can perform more sophisticated remote
operations in a single network round
trip~\cite{10.1145/3373376.3378528,li2017kv,179402,10.1145/3230543.3230572,222623,10.1145/3422604.3425923,10.1145/3342195.3387519}.
Nevertheless, such a solution 
%requires multiple round trips to the host main memory over slow PCIe links, entailing 
sometimes entails lower performance than the the first and second solutions~\cite{10.1145/3422604.3425923,neugebauer2018understanding} 
because of two reasons. 
First, accessing the server memory is inevitable and expensive when the application's data is not cached in the Smart NIC's small local memory (\secref{sec:mot}). 
This is common since modern datacenter applications typically have large working sets, often requiring not only DRAM but even high-capacity byte-addressable NVM such as Intel Optane Persistent DIMM~\cite{intel-optane} in the host memory system to cost-effectively provide necessary memory capacity~\cite{10.1145/3477132.3483565,msftnvm,googlenvm}.
Second, the Smart NIC's specialized accelerator or energy-efficient but wimpy CPU may not be powerful enough to provide the required end-to-end performance for applications. As such, it needs to participate the (beefy) server CPU for application processing, %partitioning application processing between the server CPU and Smart NIC accelerator/CPU. However, 
but it will often require frequent communications between the server CPU and the Smart NIC over slow PCIe links, which becomes a performance bottleneck.

The three strategies above pose a dilemma to system designers: whether to use expensive and precious CPU cores
for application processing, or to suffer from the performance overhead incurred by multiple round trips over network or PCIe
links. %that result from the RDMA or Smart NIC approaches.
Tackling the limitations of the current solutions, this paper proposes 
% a holistic approach that offload datacenter applications to reduce server-side CPU usage and improve energy-efficiency without the aforementioned tradeoffs: 
\arch (\underline{\textbf{O}}ffloading with \underline{\textbf{R}}DMA and \underline{\textbf{C}}c-\underline{\textbf{A}}ccelerator), a holistic modularized solution to cost-effectively and efficiently serve $\mu s$-scale datacenter applications. 
Specifically, \arch proposes a server with a standard RDMA NIC (RNIC) and a \emph{cache-coherent accelerator}
(cc-accelerator)~\cite{6138p,enzian} connected to an emerging \emph{cache-coherent off-chip interconnect} (cc-interconnect) such as CXL~\cite{cxl}. 
%Subsequently, \arch proposes a cc-accelerator architecture and a unified inter- and intra-machine communication abstraction, both of which together make the server CPU, RNIC and cc-accelerator work synergistically in a fine-grained fashion, to efficiently offload network and application processing from the server CPU.
The RNIC, cc-accelerator, and the server CPU work synergistically in a fine-grained fashion, to efficiently offload network and application processing from the server CPU.
%minimize the use of server CPU cycles for application processing.
%and (2) achieve fine-grained CPU-RNIC-accelerator co-design for end-to-end application performance.
%cc-accelerator is an emerging type of device which allows the accelerator to integrate with the CPU's cc-interconnect.
%\arch is particularly powerful because it not only allows the benefits of RDMA's network processing offload but also augments it with application-specific processing with low-overhead access to host memory and fine-grained CPU-accelerator collaboration. 
%The accelerator offers
%more predictable performance with higher power efficiency than
%pure-CPU software processing.
\arch advocates a modularized architecture (\ie, cc-accelerator as a separate device) instead of a single device integrating RNIC with a cc-accelerator (\ie, Smart RNIC), because \arch desires to reuse the standard RNIC and simply replace the cc-accelerator with one customized for other applications.
As such, \arch can  be more cost-effective or efficient for serving a wider range of applications than a solution replacing a Smart RNIC with another one integrating a different accelerator or providing sub-optimal performance with a single Smart RNIC.
To our best knowledge, \textit{\arch is the first work to explore cc-accelerator's role in various end-to-end datacenter applications}.

% In this paper, we propose \arch, a holistic end-to-end solution for accelerating memory-intensive distributed applications with RDMA and cache-coherent accelerator (cc-accelerator) technologies.
% \arch design is based on two key observations.
% First, the one-sided RDMA operations can provides the semantics of full network processing offloading -- the host only needs to deal with application request processing and can communicate with the NIC in pure userspace.
% Also, the RDMA NIC (RNIC) functionality is relatively fixed and does not require large memory footprint, making it unnecessary to be integrated to the CPU. 
% Second, the emerging cc-accelerator is a good fit for accelerating our target distributed applications with three unique advantages. 
% (1) Sharing the coherent memory space with the CPU, the cc-accelerator can access the high-capacity host memory, interact with the CPU and RNIC via the cache-coherent interconnect (cc-interconnect) with low overhead. 
% (2) The cc-interconnect is able to expose more hardware details and information under the hood to the accelerator. Leveraging such advanced knowledge, the cc-accelerator can realize more advanced fine-grained functionalities~\cite{10.1145/3445814.3446713,10.1145/3373376.3378482}. 
% (3) Compared to the pure CPU-based software solutions, an accelerator typically has more predictable performance with higher power efficiency. 
% These observations together make it possible to design and implement a high-performance, efficient, and flexible system for our purpose.

\label{sec:design}

%We depict a high-level system architecture of \arch in \figref{fig:arch}. 
\begin{comment}
%Leveraging RDMA and a cc-accelerator, \arch aims to provide 
%\arch's design principle is to make the most out of the system's heterogeneity, and offload as much work from the CPU as possible. 
%As a result, 
In \arch, the cc-accelerator is responsible for most of application processing while the standard RNIC takes care of network processing. 
%the network transport stack is offloaded to the RNIC with one-sided RDMA operations, and the cc-accelerator takes over the major routines of the target distributed applications. 
The conventional server CPU primarily plays two roles in \arch. 
%The first one is a ``controller'' or ``slow-path'' to initialize, control, and manage the hardware resources, applications, network connections, \etc. In this case, the CPU can be a low-profile energy-efficient one. 
First, the CPU initializes, control, and manage the hardware resources, applications, network connections, \etc similar to standard RDMA.
%The second is a co-processor of the cc-accelerator that deals with the irregular, branch-rich portions of the ``fast-path'' application requests. 
Second, the CPU handles the irregular, branch-rich portions of application requests as a co-processor of the cc-accelerator.
%In this case, the CPU should still be beefy to guarantee the end-to-end system performance.
\end{comment}
\arch consists of four software and hardware components that are tightly coupled and synergistically interacting with each other to cost-effectively and efficiently serve datacenter applications. Specifically, we propose:
\textbf{(1)} %we propose 
a  unified  abstraction for inter- and intra-machine communications where lockless ring buffers facilitate inter-machine communications with one-sided RDMA write and CPU-accelerator communications with \texttt{load/store};
\textbf{(2)} 
%we develop 
a fast and efficient mechanism for notification of requests to cc-accelerator exploiting the coherence information exposed to the cc-accelerators; 
\textbf{(3)} %we design 
a cc-accelerator architecture for processing the requests and handling interactions among RNIC, CPU and cc-accelerator;
and \textbf{(4)} %we introduce 
an adaptive device-to-host data transfer mechanism for %hybrid 
a server with a heterogeneous memory system consisting of 
DRAM and NVM. %. byte-addressable NVM (\eg, Intel Optane DIMM with 3D Xpoint technology).
%with both regular DRAM and emerging NVM. 
%These design components together enable the efficient end-to-end acceleration of distributed applications by leaving the CPU cores out of the critical datapath. 

\iffalse
We present the design of \arch system, which comprises four major
components t. 
Leveraging RDMA and cc-interconnect, \arch's underlying mechanism for inter-machine and CPU-accelerator communication is built atop unified lock-free ring buffer with plain memory semantics (one-sided RDMA write and coherent write, respectively) to efficiently exchange requests and responses. 
The is a new coherence-assisted
notification mechanism, called \texttt{cpoll}, that allows the
cc-accelerator to efficiently receive requests from the RNIC and CPU. Compared to regular spin-polling, \texttt{cpoll} reduces energy and avoids wasting precious bandwidth in the cc-interconnect and coherence controller.

We provide a framework for building applications atop a
cc-accelerator, allowing them to process application requests and
issue RDMA operations without the involvement of the CPU in the
critical data path.

Finally, we propose an optimized RNIC-host data transfer
mechanism that efficiently supports heterogeneous memory systems with
both DRAM and NVM. It adaptively
directs DMA requests to either the CPU's last-level cache or main
memory, avoiding a major performance bottleneck in NVM
systems~\cite{10.1145/3419111.3421294,273889}.
\fi

We prototype \arch with a commercial system based on an Intel Xeon
6138P CPU that integrates an FPGA device in the same package and communicates with the FPGA device through a UPI link.
Subsequently, we evaluate three popular $\mu s$-scale datacenter applications:
%popular in the modern datacenter: 
(1) in-memory KVS to show a case fully offloading requests and bypassing the CPU; (2)
chain replication-based distributed transaction processing system  to show a case considering a latency-sensitive system with NVM; and
(3) deep learning recommendation model (DLRM) inference to show a case processing application requests through collaboration between a server CPU and a cc-accelerator. 
We show that 
%\arch offers lower latency than the state-of-the-art solutions, while providing $\sim3\times$ better power efficiency than a conventional system based on a server CPU and standard NIC. 
\arch provides 30.1$\sim$69.1\% lower latency, up to 2.5$\times$ higher throughput, and $\sim3\times$ higher power efficiency than the current state-of-the-art solutions. 
In addition, we  demonstrate that a cc-accelerator with its local memory as part of a server's unified memory space can further improve the latency and throughput by 11.2\% and 62.1$\times$, respectively.

%%% Local Variables:
%%% mode: latex
%%% TeX-master: "paper"
%%% End:
 
% !TEX root = paper.tex
\section{Background and Motivation}
\label{sec:back}

\subsection{RDMA Primer}
RDMA is an advanced kernel-bypass network concept that allows machines to 
access the memory of remote machines at high bandwidth and low latency. RDMA has now been widely deployed by datacenters~\cite{10.1145/2934872.2934908,262036,10.1145/2785956.2787484,10.1109/TNET.2019.2961671,cx}
and used to build various research and production systems.
RDMA offloads the network transport stack and link-layer operations to RNIC hardware and supports one- and two-sided operations that can be accessed directly from the user space.
One-sided RDMA operations (\eg, read/write/atomics) completely bypass the
remote server's CPU for remote memory accesses. Meanwhile, two-sided RDMA operations (\eg, send/receive)
are similar to conventional network communications (\eg, TCP/UDP) as they involve the CPUs of both clients and servers for data transmission.
%In other words, the user only needs to prepare the memory buffer for sending/receiving data, and the NIC will handle the underlying details.
%The network transport is mostly reliable by the link layer (\eg, flow control and re-transmission) and can be in connection (all operations) or datagram (send/recv only).

The key data structures in RDMA programming are queue pair (QP) and completion queue (CQ), shared between the host (user space) and the RNIC. 
A QP consists of two work queues (WQs): a send queue (SQ) and a receive queue (RQ), both of which are ring buffers in the host memory. 
To post an RDMA operation, the user writes to a work queue entry (WQE) at the tail of the WQ with a pre-defined device-specific format and rings the RNIC's doorbell register using an MMIO write. 
Upon completion of the RDMA operation, the RNIC (optionally) writes to a completion queue entry (CQE) at the tail of the CQ (also a ring buffer in the host memory) associated with the QP. 
By polling the CQ, the host can be aware of the completion of operations.
%In general, these routines are done by calling \texttt{libverbs} APIs. 
%However, since all these structures are in the user space, the user can manually manipulate them as well.
%Note that the operations between the QP and CQ are asynchronous 
%to pipeline RDMA operations and thus maximize the hardware utilization and performance.

\iffalse
\niparagraph{The RDMA dilemma.} One-sided operations should offer the
greatest benefit with RDMA, as they bypass the remote CPU, but
building an application entirely out of read and write operations
presents a major challenge. Most systems need to perform more complex
operations in order to traverse or manipulate remote data structures,
and doing so with RDMA requires additional round trips. 
For example,
%Pilaf~\cite{180191} requires two round trips to search a key-value
%hash table: one to read the hash bucket, and a second to follow a
%pointer to the corresponding data. Similarly, 
Cell~\cite{cell}
searches a remote B-tree, potentially requiring one round trip per
level in the tree. Modifying data structures over RDMA is even more
complex, and is frequently left to a traditional CPU-based
implementation~\cite{180191,179767}. These factors make the choice of
whether to use one-sided RDMA operations or two-sided RPC a far more
complex tradeoff: the additional round trips can easily negate the
performance benefit of CPU bypass, particularly as RDMA is deployed in
larger datacenter settings with higher network latency~\cite{cx}.
\fi

\subsection{Memory Capacity vs. Communication Overhead over PCIe: Dilemma of Using Smart NIC}
\label{sec:mot}

Recent Smart NICs integrate either FPGA or customized low-profile CPU with NICs. % in order to accelerate datacenter applications.
Prior work has shown that Smart NICs running datacenter applications can  %/offloader
 offer higher performance and energy efficiency than host CPUs~\cite{li2017kv,10.1145/3477132.3483565}.  
However, Smart NICs have limited memory capacity (\textit{O(10~GB)}~\cite{10.1145/3341302.3342079}) under \textit{cost, power, and form factor} constraints. As such, they often need to access the host memory when running applications with large working sets.
%Memory accesses are a major challenge for Smart NIC offloading. The
%memory capacity of the NIC is usually limited
%, a small fraction
%of that available on the host. When applications have high memory
%capacity requirements, the Smart NIC inevitably needs to access the
%host's main memory.
Unfortunately, such host memory accesses are not cheap, primarily because they must go through PCIe links.
Specifically, the PCIe links add non-trivial latency (\eg, at least 1$\mu$s) to the access latency of the host memory and can incur performance bottlenecks~\cite{neugebauer2018understanding}, especially
when the Smart NIC needs to frequently synchronize its local memory with or retrieve data from the host memory.
This has been identified by multiple system designs~\cite{10.1145/3477132.3483565,li2017kv,neugebauer2018understanding,10.1145/3477132.3483555,10.1145/3477132.3483587}, and we also observe the same phenomenon on a BlueField-2 Smart NIC; for a Smart NIC application accessing both its local on-board memory and host memory, the latency increases and throughput decreases linearly with higher percentage of memory accesses to the host.
Consequently, the Smart NIC is only well suited for applications with 
working sets that are either small enough to fit in the Smart NIC's local on-board
memory or effective for caching (\ie, sufficient spatial locality or skewed distributions of memory accesses).
%Consequently, the %compute 

%In summary, the Smart NIC 
%Otherwise, the frequent NIC-host communications will adversely affect the end-to-end performance of applications.

\subsection{Cache-coherent Interconnects and Accelerators} % Attached to Cache-coherent Interconnects}
Originally, cc-interconnects were developed for NUMA systems where CPUs share the memory space in a cache coherent manner.
Recent such cc-interconnects include UPI~\cite{intel-upi}, Infinity Fabric~\cite{amd-if}, and ECI~\cite{eci}. % is natively cache-coherent as part of the CPU's coherence system.
Lately, with emergence of accelerators and need for fine-grained data sharing between CPUs and accelerators, diverse cc-interconnects such as CXL~\cite{cxl}, CAPI~\cite{10.1147/JRD.2014.2380198}, and CCIX~\cite{ccix} built atop standard chip-to-chip physical links such as PCIe have been proposed.
Accelerators built on such cc-interconnects are referred to as cc-accelerators. 
As cc-interconnects facilitate cheaper host-accelerator communications for sharing a small amount of data, cc-accelerators can be more efficient than conventional accelerators connected to PCIe.
This makes cc-accelerators suitable for $\mu$s-scale acceleration/offloading. 
Currently, some cc-accelerators are commercially available~\cite{6138p,enzian,10.1145/3294054} and more commercial cc-accelerators will emerge as the next-generation Intel Xeon CPUs begins to support CXL~\cite{sapphire}.

\section{\arch System Architecture}
%\arch is a holistic solution consisting of design aspects from communication and interaction between and insides machines, accelerator architecture, as well as data transfer optimization between devices and the (heterogeneous) host memory hierarchy. 

\begin{figure}[!t]
    \centering 
    \includegraphics[width=\linewidth]{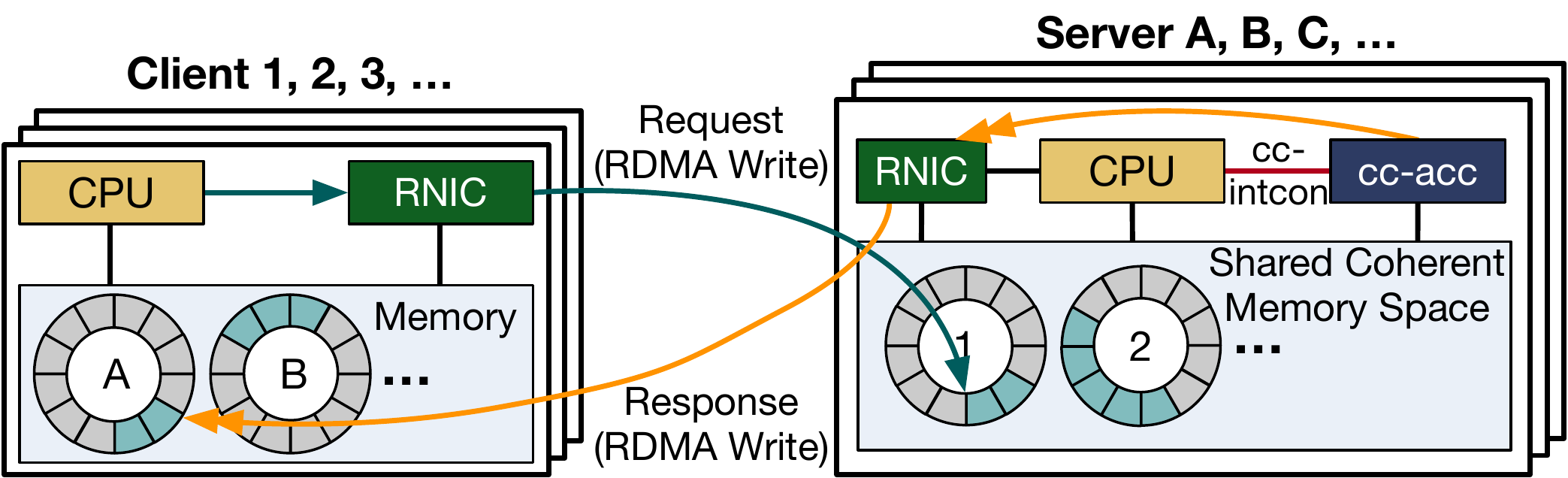}
    \caption{\arch's high-level system architecture; %Clients are indexed by ``1, 2, 3'' and servers are indexed by ``A, B, C''. 
    only the snapshot of client-1 and server-A are demonstrated.}
    \label{fig:arch}
\end{figure}

We depict a high-level system architecture of \arch in \figref{fig:arch}. 
%Leveraging RDMA and a cc-accelerator, \arch aims to provide 
\arch envisions a system comprising an emerging cc-accelerator, a standard RNIC and a conventional powerful server CPU, each of which plays indispensable roles for efficient network and application processing.
% design goal is to make %the most out of the system's heterogeneity by offloading as much work from the server CPU as possible. 
%Toward this goal, we architect \arch such that 
Specifically, (1) the cc-accelerator not only offloads accelerator-friendly part of application processing from the server CPU but also directly communicates with clients through the RNIC, \ie, receiving requests and sending responses from/to clients without involving the server CPU; (2) the standard RNIC handles network processing; and (3) the server CPU tackles accelerator-unfriendly (irregular and branch-rich) part of application processing %(``fast path'') 
in addition to initialization, control, and management of hardware resources, applications and network connections. % (``slow path''). 
The synergistic orchestrations among (1) -- (3) are facilitated by the \arch's four software and hardware components described in this section.

\subsection{Inter- and Intra-machine Communication}
\label{sec:communication}

\arch proposes a communication abstraction to (1) accomplish fast and efficient communications not only between a server and clients (\ie, inter-machine communications) but also between a CPU and a cc-accelerator in a server (\ie, intra-machine communications) and (2) unify the programming model for both inter- and intra-machine communications, based on lock-free ring buffers. %\arch cc-accelerator, . % for communications between a server and clients as well as between a CPU and a cc-accelerator in a server.  
%As illustrated in \figref{fig:arch}, for each client-server connection (\ie, each RDMA connection or CPU-accelerator pair), we have one request buffer and one response buffer. 
%As illustrated in \figref{fig:arch}, 
For each client-server connection we establish a pair of a request ring buffer (in the server memory) and a response ring buffer (in the client memory) for inter-machine RDMA communications. For example, buffer-1 on server-A and buffer-A on client-1 form a request-response buffer pair for a connection between client-1 and server-A in \figref{fig:arch}. 
Besides, for each accelerator in a server, we establish one request-response ring buffer pair in the server memory for
intra-machine communications.
One-sided RDMA write is used by both servers and clients for high-performance inter-machine communications through message passing
where all the underlying network transport processing is offloaded to the RNIC~\cite{179767,10.1145/3373376.3378528}.
For the intra-machine communications, leveraging the shared coherence domain, the server CPU or cc-accelerator directly writes to or reads from the ring buffers.

Note that we do not share the ring buffers (and the underlying RDMA QPs for the inter-machine communications) across different client-server connections, to avoid performance overheads with atomic updates or consistency issues at the head/tail of the buffer without atomic updates. 
However, we do allow sharing the ring buffers (and the RDMA QPs) across threads on the same machine for better scalability, as a software layer/library can manage cross-thread contentions with slight performance overheads~\cite{10.1145/3477132.3483576,10.1145/3302424.3303968,10.1145/3132747.3132762}.
Specifically, we employs the Flock's method~\cite{10.1145/3477132.3483576}, \ie, one dedicated thread on the client for request synchronization and dispatch, so that there is only one request-response buffer pair (and QP) per client-server pair per application and observe no performance loss compared to native RDMA primitives.

%Similar to prior work~\cite{179767,10.1145/3373376.3378528}, we use a pair of ring buffers to pass messages across machines. 

\iffalse
As illustrated in \figref{fig:arch}, for each RDMA connection (\eg, each client-server pair), we have one request buffer on the server and one response buffer on the client. 
One-sided RDMA write is used for message passing.
That is, the host directly processes the transmitted data as all the underlying transport is offloaded to the NIC.
We do not share buffers (and the underlying QPs) across connections, in order to avoid consistency issues
at the head/tail of the buffer or require expensive atomic updates. 
However, we do allow buffer (QP) sharing across threads on the same machine for better scalability, as a software layer/library can manage cross-thread contentions with minimal overhead~\cite{10.1145/3477132.3483576,10.1145/3302424.3303968,10.1145/3132747.3132762}.
Specifically, we employs Flock's method~\cite{10.1145/3477132.3483576} (in short, one dedicated thread on the client for request synchronization and dispatch) so that there is only one buffer pair (and QP) per client-server pair per application and observe no performance loss compared to native RDMA primitives.
\fi

The client is responsible for tracking the tail of the request buffer in the server memory and the head of the response buffer in its local memory, similar to the credit-based flow control~\cite{10.1145/190314.190324}.
Whenever it writes a message to the request buffer, it will update its local record of the request buffer's tail;
 whenever it receives a message in the response buffer (by polling), it will update its local record of the response buffer's head and reset the buffer entry to ``0''.
Only if the request buffer's tail is behind the response buffer's head can the client issue a request. 
Otherwise, it knows that the buffer is full of on-the-fly requests and should not send more requests. 
A similar mechanism is applied to the server for request buffer's head and response buffer's tail.
This guarantees that any message can be passed by only one network trip without any conflict.

%\iffalse

%\fi

\subsection{Coherence-assisted Accelerator Notification}
\label{sec:cpoll}

Since a client and a server CPU directly write messages to inter- and intra-machine communication request buffers in the server memory, respectively, the \arch cc-accelerator needs to proactively get the message from the request buffers.
Typically, a spin-polling mechanism can be deployed. 
However, the bandwidth of the cc-interconnect and the coherence controller is precious.
Frequently polling the request buffers, the cc-accelerator has little bandwidth left for application processing, \ie, accessing the server's memory to serve requests. % from memory-intensive applications;
%if the polling interval is long, the application-level latency of the request will be affected.
Besides, polling has high power consumption~\cite{10.1145/3357223.3362737,196292,4404806,10.1145/3225058.3225129}, fast transistor aging~\cite{6263957}, and poor scalability with many queues~\cite{10.1145/3357223.3362737}.
Hence, we propose a coherence-assisted notification mechanism, called \texttt{cpoll}.

\begin{figure}[!t]
    \centering 
    \includegraphics[width=\linewidth]{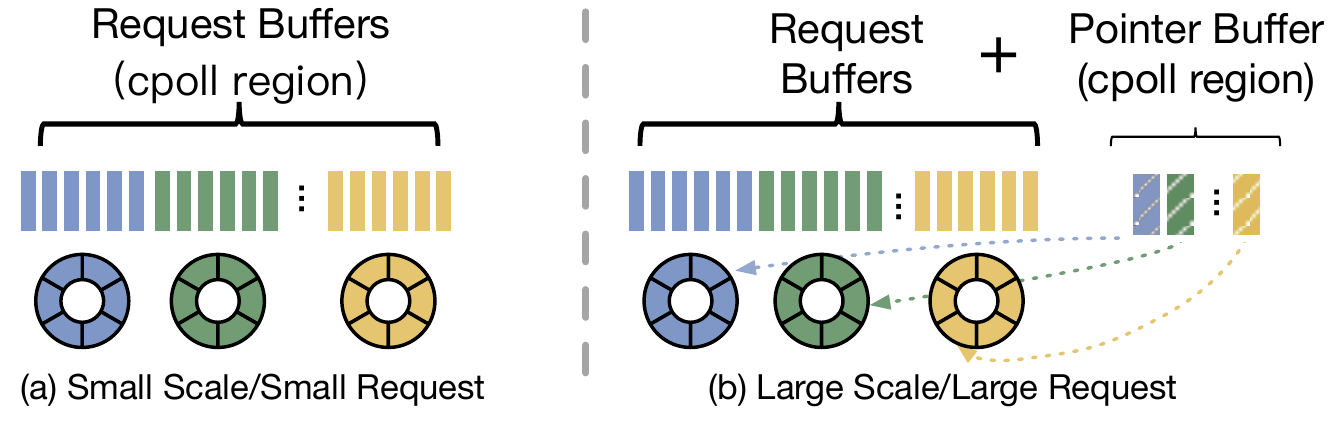}
    \caption{A \texttt{cpoll} region for cc-accelerator notification.}
    \label{fig:cpoll}
\end{figure}

Conceptually and semantically, \texttt{cpoll} is similar to \texttt{MWAIT} in the x86 architecture~\cite{sdm2}, \texttt{QWAIT} in HyperPlane~\cite{hyperplane}, and PCIe's lightweight notification (LN) proposal~\cite{ln-amd,ln-plda}. 
Nonetheless, \texttt{cpoll} differs from them because it is designed to be portable and platform/CPU-agnostic for off-chip devices. 
Specifically, we insert a \texttt{cpoll checker} in the datapath of the coherence controller's port connected to the cc-interconnect. 
During initialization, we first allocate the inter- and intra-machine communication request buffers (\secref{sec:communication}) in a contiguous address region of the server memory (\ie, \texttt{cpoll} region), and register this region to the cc-accelerator's \texttt{cpoll checker} for snooping, as depicted in \figref{fig:cpoll}(a). 
Then, when the cc-accelerator's coherence controller receives a coherence signal from the registered address region (\eg, \texttt{Modified} $\rightarrow$ \texttt{Invalid}), it will notify the cc-accelerator of arrival of a request. The \texttt{cpoll checker} will only need to monitor a single address region illustrated in \figref{fig:cpoll}(a).
%, \ie, whether lower\_bound $\leq$ address of incoming coherence signal $\leq$ upper\_bound. 
If the address of a coherence signal falls into this region, the \texttt{cpoll checker} can identify which request buffer (associated with a specific client or the server CPU) that received a new request by determining the address offset from the %lower\_bound 
starting address of the region as the size of buffers is fixed after the initialization. Hence, there will be no scalability concern for the \texttt{cpoll} mechanism. Even if the buffers are not allocated consecutively in the memory, the overhead of address lookup should not be a concern as well (to \textit{O(1K)} level buffers at least), as demonstrated by HyperPlane~\cite{hyperplane}. 
%Also note that the \texttt{cpoll} mechanism does not generate extra traffic on the cc-interconnect since the request data needs to be consumed by the cc-accelerator anyway -- \texttt{cpoll} only monitors and selects signals from the existing coherence traffic. 

%To make it portable and platform/CPU-agnostic, \textit{\texttt{cpoll} should not involve CPU-side changes or depend on any specific protocols}. 
%Keeping \texttt{cpoll}'s design principle in mind, 
To implement the \texttt{cpoll} mechanism, we propose two approaches. 
First, we allocate the \texttt{cpoll} region to the memory attached to the CPU, and then pin the region on the cc-accelerator's local cache. 
Note that the cc-accelerator resets the request buffer entry associated with a \texttt{cpoll} signal after it completes processing the request. This makes the cc-accelerator's local cache always own the \texttt{cpoll} region from the cache coherence viewpoint, and any change to the \texttt{cpoll} region by clients or the server CPU trigger a coherence signal.
%, to be ready for the next round of requests as a ring buffer, the request buffer's entries are reset by the cc-accelerator after the \texttt{cpoll} signal is processed. 
%This guarantees that the cc-accelerator's local cache always has the ownership of the data, and that any change of the 
%\texttt{cpoll} region will trigger a coherence signal. 
Alternatively, we allocate the \texttt{cpoll} region to the memory attached to the cc-accelerator.
As such, any request from the RNIC to the request buffers in the server memory space (consisting of CPU and cc-accelerator memory regions) will go through the cc-interconnect. Subsequently, they will be delivered to the cc-accelerator's coherence controller that is responsible for monitoring any change to the \texttt{cpoll} region.

The first approach is feasible with our prototype platform (\secref{sec:prototype}), but the size of request buffers is constrained by the cc-accelerator's local cache size, limiting its scalability at the moment. When the scale of the system is large (\ie, many request buffers) or each request itself is large (\ie, large buffers), we cannot pin the entire \texttt{cpoll} region on the cc-accelerator's cache. 
%The second approach less likely encounters this scalability issue, but requires a cc-accelerator with its local memory~\cite{enzian,samsungcxl,stratix10dx}, which is not at our disposal. 
%Hence, we take the first approach in this paper. %, and leave the second approach in the future. 
To tackle this scalability issue in our setup, we introduce a data structure called pointer buffer where each 4-byte entry corresponds to each inter- or intra-machine request buffer and stores a pointer (or index) to an entry in the request buffer, as depicted in \figref{fig:cpoll}(b). Subsequently, we register the pointer buffer allocated to a contiguous address space as the \texttt{cpoll} region. 
When writing a new request to a request buffer in the server memory, a client or the server CPU will also increment the value of the pointer buffer entry corresponding to the request buffer such that the pointer buffer entry points to 
%the index of 
the request buffer tail.
%each request buffer entry can be as large as a few KB depending on applications 
%Each request buffer will have a 4-byte entry back-to-back in the counter buffer, and only the counter buffer is registered as \texttt{cpoll} region. 
For a remote client, this can be efficiently done by posting two contiguous WQEs (only the second one is signaled) with a batched doorbell to the RNIC~\cite{196243} or remapping/interleaving the two buffers with user-mode memory registration (UMR)~\cite{rdmamanual} and only posting one WQE. 

Note that one additional small PCIe write to the server side is inevitable in both ways. 
However, since \arch has already reduced the PCIe traffic and mainly leverages the coherence traffic, such overhead will not notably hurt the overall performance, which is confirmed by our experiments  in \secref{sec:evaluation}. %(we do not observe performance drop by the counter buffer in our experiments). 
In addition, as a 4-byte pointer buffer entry covers an entire request buffer, which can be as large as several MBs for some applications such as the one described in \secref{sec:trans}, it can substantially reduce the memory space requirement for the \texttt{cpoll} region. %For example, ....
Finally, coherence signals are not guaranteed to come in the order of actual data writes. However, this does not affect the correctness of \texttt{cpoll} because it is designed to be used with a ring buffer, and leverages the semantics of the ring buffer, \ie, request buffer entries are written in order. %Hence, if the \texttt{cpoll} signal of an entry (or a pointer value) is received by the cc-accelerator, all entries before it have been received and can safely be processed. 

\begin{figure}[!t]
    \centering 
        \includegraphics[width=\linewidth]{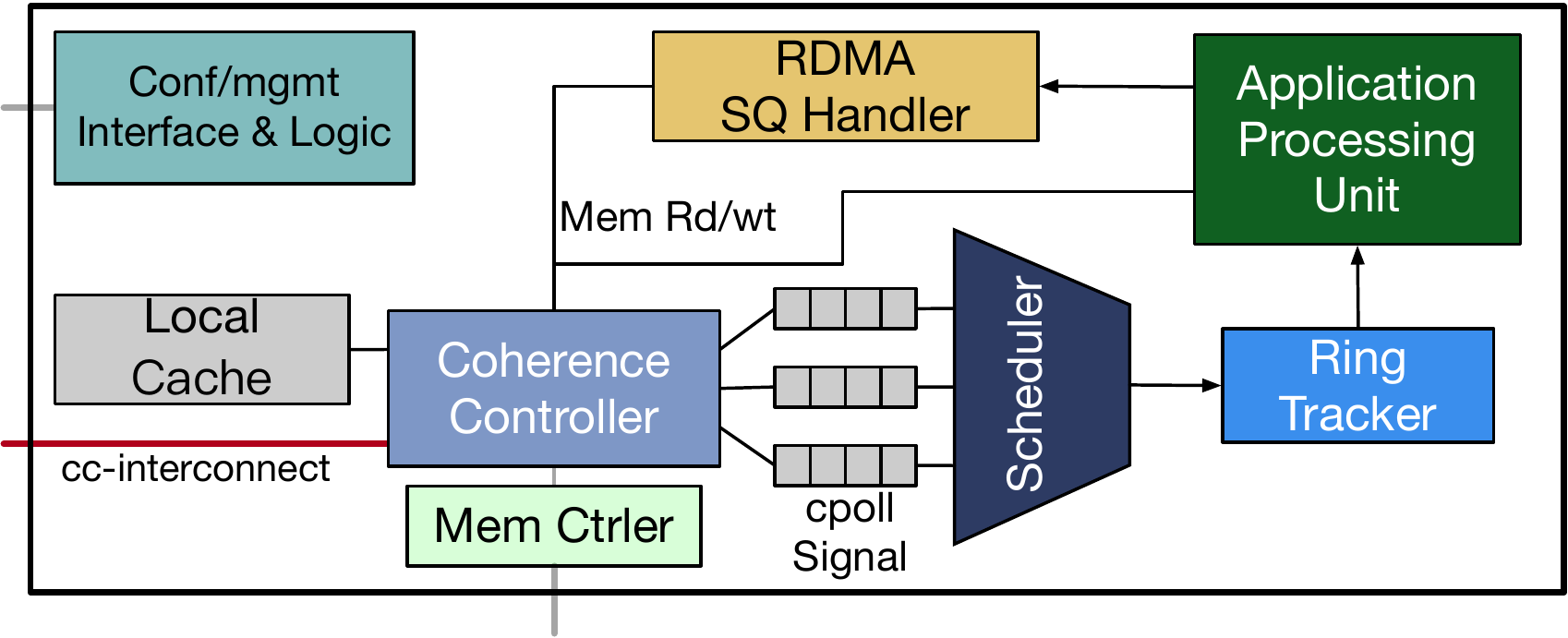}
    \caption{\arch cc-accelerator architecture.}
    \label{fig:accel}
\end{figure}
\subsection{\arch cc-accelerator Architecture}
\label{sec:accel}
We depict \arch cc-accelerator architecture in \figref{fig:accel}. 
The coherence controller handles all the coherence traffic (both regular read/write and \texttt{cpoll}) to/from the cc-accelerator, as well as the virtual-physical address translation (\ie, TLB).
The local cache is also in the coherence domain and handled by the coherence controller.
%The registered \texttt{cpoll} region is pinned in the local cache.
The \arch cc-accelerator may have its own local memory controller and memory~\cite{samsungcxl,enzian} that constitutes the unified memory space with the CPU memory. 
In such an architecture, the CPU may allocate application data to the cc-accelerator's local memory, as the NUMA-aware memory management does in the modern Linux kernel.
%This will become possible in the near future.

The scheduler fetches \texttt{cpoll} signals associated with different request buffers based on a given  scheduling algorithm. % for QoS purpose.
Due to the nature of the coherence implementation, \texttt{cpoll} signals can be coalesced. 
For example, if we update the same entry in the pointer buffer twice in a short period, there can be only one \texttt{cpoll} signal generated. 
However, leveraging the semantics of the ring buffer, \ie, a pointer value only increments (including \textit{mod}), we introduce a ring tracker to the cc-accelerator to track the previous tail of the request buffer. 
It tells the application processing unit (APU) how many new requests are received since the last notification based on the difference between the recorded tail pointer value and the incoming pointer value.

The RDMA SQ handler is responsible for assembling the response information into the format of the RNIC's WQE and ring the RNIC's doorbell register in its PCIe BAR. 
Since polling the CQ is not on the critical datapath, we do not process it with the cc-accelerator. Instead, we use a single CPU core to handle all the CQs polling and bookkeeping.
%As only the selected operations will notify the CQ of their completion, unsignaled WQE~\cite{196243} %(\ie, only the selected operations will notify the CQ of their completion) 
%is applied here to alleviate the overhead of RNIC-CPU communication when the CPU is polling multiple CQs.
Unsignaled WQE~\cite{196243} is applied here so that only the selected operations will notify the CQ of their completion.
This can alleviate the overhead of RNIC-CPU communication when the CPU is polling multiple CQs.
Besides, this helps reduce unnecessary traffic on the cc-interconnect.

The APU is \textit{the only application-specific part} in the entire \arch architecture, yielding a fine balance between \arch programmability and user implementation complexity. 
It provides the user with \textit{standard interfaces} for (1) \texttt{cpoll} signal reception, (2) coherent data read/write, and (3) RDMA WQE output.
%The application processing unit receives \texttt{cpoll} signals and process requests. After it produces the result, it will notify the RDMA SQ handler of sending response. 
%Although the request processing logic differs from application to application, many applications will have a similar high-level organization, allowing \arch components to be reused.
First, a (de)serializer can be optionally used, if the application uses an RPC protocol for inter-machine communications~\cite{10.1145/3445814.3446696}. 
Then, to process requests, we typically need a data structure walker~\cite{9407097,kocberber2013meet,10.1145/3373376.3378497,9407194,10.1145/3079856.3080234} to find the location of the target data of the request.
To maximize the memory-level parallelism and hide the memory access latency, multiple outstanding requests and out-of-order execution should be supported.
Inspired by the stateful network function accelerator~\cite{225996}, we employ a table-based finite state machine for this purpose, where the outstanding request status is stored in a TCAM or cuckoo hash table~\cite{zhou2013scalable} for fast lookup. Upon the arrival of a new request or intermediate result, the corresponding TCAM or hash table entry is updated and then the next-step action is issued to a corresponding functional unit (\eg, ALU or coherence controller). 

The APU should invoke the CPU in two scenarios. 
The first scenario is when a library call or OS \texttt{syscall} is needed. 
For example, if the user space memory pool has been pre-allocated by the CPU (\texttt{malloc/mmap}), the APU itself can allocate objects for new data in the memory pool~\cite{li2017kv}; if not, \texttt{malloc} is called each time when a new object is needed.
The second scenario is when CPU is more suitable than the APU for a certain part of application processing. 
For example, in a recommender inference system (see \secref{sec:recom}), while the APU can handle the embedding reduction and fully-connected layers, the request preprocessing (\eg, transforming a human-readable request to a model input) should still run on the CPU due to its irregularity and complexity. 
In these scenarios, the cc-accelerator and CPU interact with low latency in a fine-grained manner described in \secref{sec:communication}.

\iffalse
, we advocate for CPU-based software allocation (\ie, \arch cc-accelerator invokes the CPU core to call \texttt{malloc} functions) for three reasons. 
(1) Hardware-based memory allocator (\eg, slab allocator~\cite{li2017kv}) is not as flexible and efficient as the various software-based allocators~\cite{tcmalloc,jemalloc} when it needs to switch between the kernel and userspace. 
(2) The expensive part of the memory allocation procedure can be alleviated by the near-core memory allocation accelerators~\cite{kanev2017mallacc}. 
(3) The CPU shares the coherence domain with the \arch cc-accelerator, meaning the overhead of invoking CPU-based \texttt{malloc} is low. 
\fi

%All these components are initialized and controlled by the CPU via a configuration/management interface (with corresponding logic on the accelerator) at runtime.

\subsection{Optimizing Device-host Data Transfer: Adaptive DDIO}
\label{sec:ddio}
Having the \arch system design, we finally consider the optimization of device-memory-cache interaction inside a single machine, or specifically, how to choose between the cache and memory as the data destination for optimal device-host data transfer.
Given the data-intensive nature of \arch's usage scenarios, this optimization is notably important for the entire system.
As the device I/O speeds increase, 
Intel introduced data-direct I/O (DDIO)~\cite{ddio}, a CPU-wide technology, to allow the device to directly inject data to the CPU's last level cache (LLC) instead of main memory. This reduces memory bandwidth consumption and latency required by I/O.

%Attracting much attention in the architecture and system communities~\cite{alian-ispass,10.1145/3419111.3421294,tootoonchian2018resq,kurth_netcat:_2020,254372,farshin2019make,yuan-iat,273889,100glinux,10.1145/3387514.3405868}, 
DDIO has been proven to be effective~\cite{alian-ispass,10.1145/3419111.3421294,tootoonchian2018resq,kurth_netcat:_2020,254372,farshin2019make,yuan-iat,273889,100glinux,10.1145/3387514.3405868,10.1145/3503222.3507711, nfslicer}, improving the performance of DRAM-based systems~\cite{alian-ispass,kurth_netcat:_2020,100glinux}. However, it does not always improve performance of NVM-based systems~\cite{10.1145/3419111.3421294,273889}. which is increasingly deployed by datacenters to cost-effectively provide large memory capacity for applications such as in-memory database~\cite{msftnvm,googlenvm}.
This is mainly because of two reasons. 
\textbf{(1)} NVM has a larger access granularity than DRAM and cache. For example, the access granularity of the Intel Optane DIMM is 256 bytes while that of DRAM and cache is 64 bytes in Intel-based system~\cite{246192}.
When the DDIO-ed data is evicted from LLC to NVM, %(due to LLC capacity limit), \
the write-back to the NVM will be randomized because of the cache replacement policies. 
As a result, write amplification wastes the bandwidth of NVM~\cite{10.1145/3419111.3421294}, which is already lower than that of DRAM.
\textbf{(2)} CPU caches are typically not persistent; Intel eADR~\cite{eadr} makes cache as part of the persistency domain but it requires a large battery and has high power consumption~\cite{9407087,eadr-cost}.
%\footnote{With , cache can be included into the persistency domain, but it requires a large battery and has high power consumption~\cite{9407087,eadr-cost}.}
As such, applications often have to flush data in cache to NVM to remain correct in case of a crash, with performance cost~\cite{rdma-pmem}.
%This may affect persistency requirements (and the performance) of the applications. 
%For instance, to write data to the remote NVM, an RDMA write must be followed by another RDMA read to guarantee that the data has been flushed to NVM~\cite{rdma-pmem}. 
%In summary, if a server co-runs DRAM- and NVM-based applications, the performance of one application is always negatively affected, depending on whether we enable . 

\begin{figure}[!b]
    \centering 
    \includegraphics[width=\linewidth]{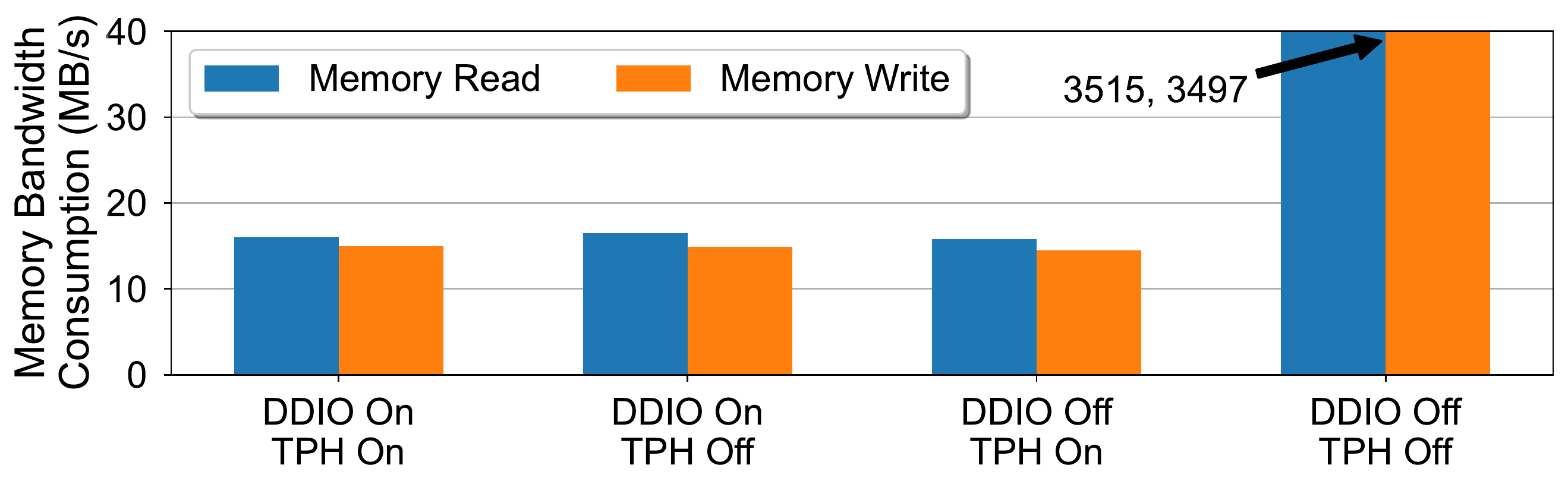}
    \caption{Memory bandwidth consumption by PCIe-bench's DMA write with different DDIO/TPH settings.}
    \label{fig:tph}
\end{figure}

%Can we control DDIO in a finer granularity (or achieve the equivalent behavior)? 
To tackle the aforementioned limitation, % of the default DDIO, 
we propose to exploit a rarely-discussed field in the PCIe packet header, TLP processing hints (TPH). % bit -- can be leveraged to achieve this goal. 
It is the $16^{th}$ bit in the PCIe header and a performance feature that allows the CPU to prefetch or keep certain PCIe writeback data in LLC for quick consumption by CPU cores~\cite{tph}. 
To our best knowledge, no current commercial I/O device (including SSD and NIC) uses the TPH bit; it is always set to 0 as a placeholder in both hardware and device drivers.

Our experiment confirms that changing the TPH bit allows us to control the destination of data to either LLC or memory per PCIe packet.
%, on a per-PCIe-packet level, whether data is written to the LLC or to memory. 
\textit{Being the first to clarify the DDIO-TPH relationship with modern server platforms}, we perform an experiment running PCIe-bench~\cite{neugebauer2018understanding} on a Xilinx VC709 FPGA board~\cite{vc709} in which we implement a module that allows an API to set the TPH bit on-the-fly for each PCIe packet.
The FPGA DMAs (random write) data to the (DRAM-based) host at a constant speed of 3.5GB/s. 
We measure the memory bandwidth consumption on the host side % to indicate the destination of the DMA-ed data -- if the data is sent to LLC, the memory bandwidth consumption should be low and vice versa. 
%We measure the value
in four configurations (DDIO on/off + TPH on/off) in \figref{fig:tph}.
Only when both DDIO and TPH are off, we observe large memory bandwidth consumption (\ie, $\sim$3.5GB/s for both read and write), which is aligned with the DMA throughput reported by the FPGA. This indicates that all DMA data is sent to the main memory.
Otherwise, if \emph{either}  DDIO or TPH is on, there will be little memory bandwidth consumption, meaning data is sent to the LLC directly.

\begin{figure}[!t]
    \centering 
    \includegraphics[width=\linewidth]{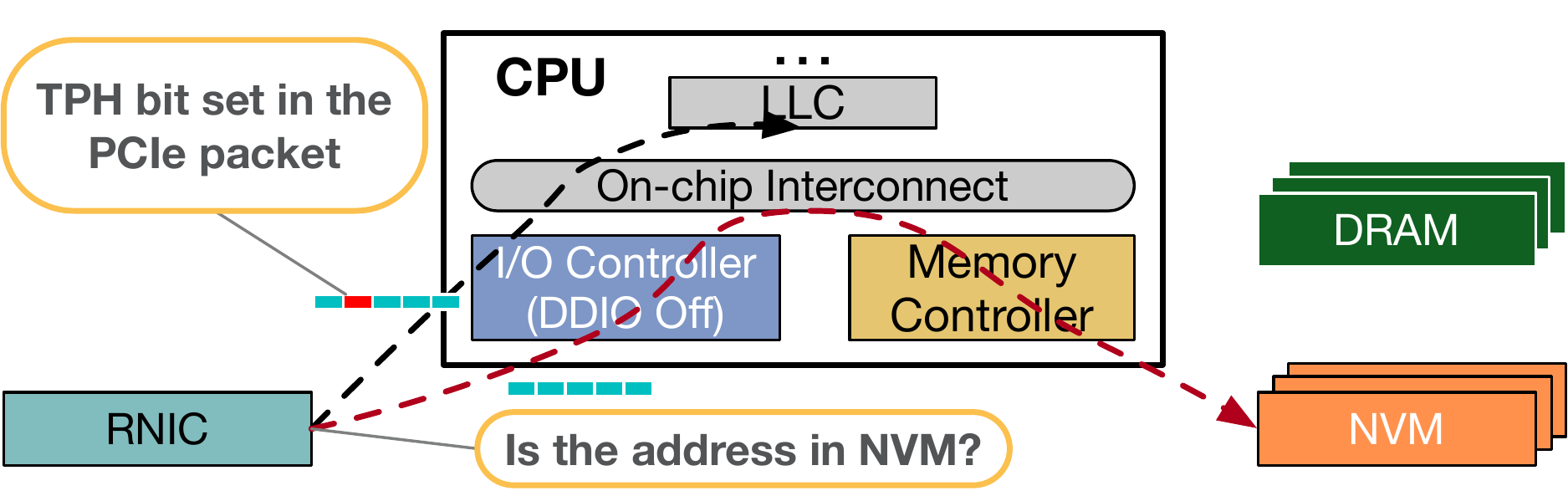}
    %\vspace{-6pt} 
    \caption{DDIO/TPH configurations in the system with NVM.}
    \label{fig:ddio}
\end{figure}

Since TPH is applicable to each PCIe packet, we propose two guidelines for future systems with heterogeneous memory, as depicted in \figref{fig:ddio}. 
(1) DDIO should be disabled globally on the CPU by default. 
(2) The device should expose the knob of changing the TPH bit for the programmer. 
Taking RNIC as an example, one way to do this would be to make it a configuration parameter set when registering a memory region to the RNIC, specifying whether the registered address range belongs to DRAM or NVM. 
Later, when executing an RDMA operation (\eg, write), the RNIC hardware will set the TPH bit only for operations in DRAM regions, avoiding DDIO-induced write amplification on NVM regions. 
While this requires hardware modifications to the RNIC, our experiment that adds this functionality to the original PCIe-bench shows that it adds almost no cost to the NIC hardware design.
Following these two guidelines, we can make DDIO  NVM-aware independently for each I/O device.

\section{\arch Use Cases}
\label{sec:usecase}

\subsection{In-Memory Key-Value Store}
\label{sec:kvs}
In-memory key-value store (KVS) is a basic building block of many datacenter services. % distributed systems and infrastructures. 
In a KVS, the key-value pairs are usually organized in a hash table or B-tree for fast lookup.
In this paper, we take the hash table based KVS as a use case. 
Upon arrival of a request, a hash value is calculated based on the requested key.
Based on the hash value, a specific hash table entry/bucket is accessed for data retrieval/update/insert. 
To avoid hash collisions, chaining~\cite{179747} or cuckoo hashing~\cite{fan2013memc3} can be leveraged, both of which often increase the number of memory accesses. 
In addition to software optimizations, researchers have leveraged all the three directions mentioned in \secref{sec:intro} and \tabref{tab:comp} for further KVS acceleration. 
The major requirements of KVS is \textit{high memory access parallelism across requests}.

\begin{comment} 
First, regular kernel/userspace network stack can be replaced by (two-sided) RDMA for fast and low-overhead message passing~\cite{225980,196243,10.1145/2806887,10.1145/2619239.2626299}, while the server CPU still needs to process requests and can become the bottleneck. 
Second, the compute/request processing can be offloaded to the client and (one-sided) RDMA is just needed to retrieve data from the remote server, which requires multiple network round-trips~\cite{180191}.
This approach can be further optimized by offloading the request processing to the server's Smart NIC~\cite{li2017kv,10.1145/3422604.3425923,10.1145/3342195.3387519} so that only one one-sided RDMA operation is required per request. 
\end{comment}

%\subsection{\arch KV Design}
%\niparagraph{\arch KV design.}
An \arch design for KVS, dubbed \arch KV aims to \textit{fully offload} %``fast-path'' 
request processing to the cc-accelerator.
At the algorithm/data structure level, \arch KV is similar to MICA~\cite{179747}, but \arch KV follows the architectural description of \secref{sec:accel} at the hardware level, including a pipelined hash unit for hash value/index calculation.
\arch KV performs a GET/UPDATE request by calculating the hashed key value and finding the corresponding entry in the set-associated hash table's bucket. The entry contains a pointer to the actual key-value data. For PUT requests, after finding the address where a new key-value pair should be allocated (\ie, an empty entry in the bucket), the slab allocator will simply put it in the pre-defined memory pool. If the bucket indexed by the hashed key is full (\ie, hash collision), another bucket with the same format will be allocated and linked to the existing bucket by a pointer. 
Similar to KV-Direct~\cite{li2017kv} and MICA~\cite{179747}'s study, on average, each GET request requires three memory accesses and each PUT request requires four.

\subsection{Distributed Transaction with NVM-based Chain Replication}
\label{sec:trans}

Distributed transactional systems are widely used by datacenters to provide the ACID feature for distributed storage systems.
To this end, cross-machine protocols for data replication is usually needed, and chain replication~\cite{mongochain,sapchain,10.1145/2465351.2465361,10.1145/1629575.1629577,180277,10.1145/2043556.2043571,10.1145/2342356.2342360,10.1145/945445.945450,10.1145/3064176.3064215,10.1145/2043556.2043560,10.1145/2378356.2378361,196235,267684,10.5555/1251254.1251261,265017} is a popular primary-backup replication protocol.
In chain replication, machines are virtually organized into a linear chain. 
Any change to the data will begin at the head of the chain and pass through the chain. 
When the last machine in the chain makes the change in its log, it will back-propagate the ACK signal through the chain so that each machine can locally commit the transaction. 
When the head of the chain commits the transaction, it sends the ACK signal back to the client, marking the end of the transaction.

\begin{figure}[!t]
    \centering 
    \includegraphics[width=\linewidth]{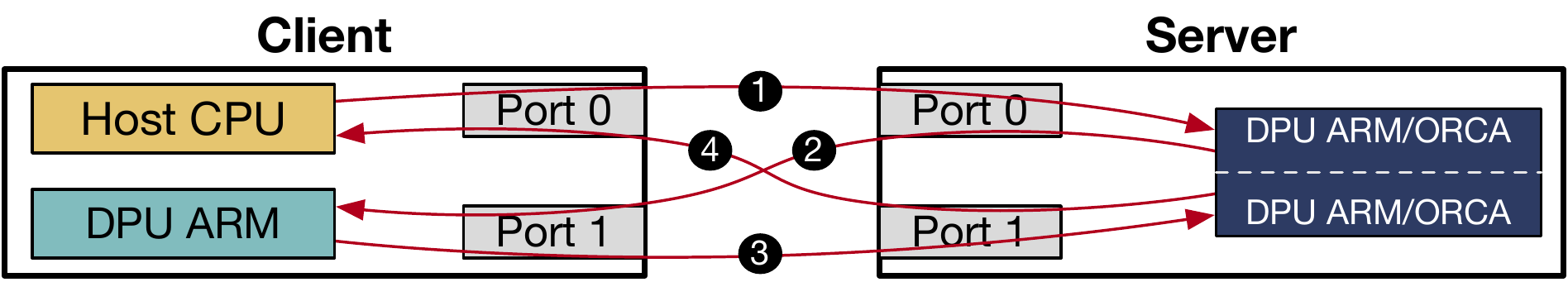}
    %\vspace{-1ex}
    \caption{Emulated 2-node replication; the DPU ARM in the client simply forwards data from port 0 to port 1, and the DPU ARM/\arch accelerator is separated to handle traffic from port 0 and 1 independently.}
    \label{fig:chain}
\end{figure}

The state-of-the-art work, HyperLoop~\cite{10.1145/3230543.3230572}, leverages the RNIC and NVM to achieve low-latency chain replication with little CPU involvement.
Specifically, it proposes and implements group-based RDMA primitives, which can be triggered automatically by the RNIC. 
One key-value pair (addressed by the offset in the NVM space) is modified in the entire chain once the client initiates a group-based RDMA operation. 
However, to process multi-value transactions, the client needs to sequentially issue RDMA operations for each key-value pair, which often leads to long latency in the network and PCIe link.

%\niparagraph{\arch T design.}
An \arch design for such a distributed transaction system, dubbed \arch TX is similar to \arch KV with respect to request processing, but it additionally implement a concurrency control unit in the APU. That is, any single key-value pair can only be accessed by one outstanding transaction, and the other related transactions will be buffered in the queue in the order of arrival.
The concurrency control unit is a small hash table, and its entries are indexed by the key of the key-value pair. 
Key-value pairs are stored in the NVM and accessed by the address offset relative to the starting address, which is the same as HyperLoop. 
Also, it adds the functionality of chain-based communication across replica machines. 
The inter-machine communications still rely on ring buffers described in Sec.~\ref{sec:communication}, but the ring buffers are allocated in the NVM as the redo-log for failure recovery.
One log entry (transaction) can contain multiple \texttt{(data, len, offset)} tuples, and the first byte of the log entry indicates the number of tuples.
One exception is pure read transactions. 
Similar to HyperLoop, since the chain replication protocol already provides data consistency, a client can conduct a pure read transaction by directly accessing the chain's head/tail machine with one-sided RDMA read.

\subsection{DLRM Inference}
\label{sec:recom}
DLRMs have received much attention by Internet giants~\cite{10.1109/ISCA45697.2020.00084,9065589,10.1109/ISCA45697.2020.00070,9407135,naumov2019deep,8259424,10.1145/3394486.3403341} as they can offer more revenue and better user experience.
In an end-to-end recommender system, the most expensive part of serving an inference request is the embedding reduction step, consuming huge memory capacity~\cite{naumov2019deep,zhao2020distributed,10.1145/3357384.3358045} and $\frac{1}{2}\sim\frac{3}{4}$ of the inference time~\cite{10.1145/3445814.3446717,eisenmanbandana,9065589,10.1109/ISCA45697.2020.00070,10.1109/ISCA45697.2020.00083}.
 %\textit{taking up huge memory capacity} to be stored
The embedding reduction operation processes queries on a set of features. 
It finds a (sparse) embedding vector in the embedding table (a high-dimension matrix) and aggregates a value.
The values of all features are assembled as the result.
Also, the embedding reduction is bounded by memory bandwidth and exhibits poor data locality~\cite{10.1145/3445814.3446717,10.1109/ISCA45697.2020.00070}.
Last but not least, it also incorporates routines like request parsing and transforming (pre-processing), which are irregular and branch-rich. As such it is not suitable for hardware accelerators.  
These characteristics make it unsuitable, if not impossible, to be fully offloaded to any Smart NIC or cc-accelerator.
In addition to acceleration with specialized hardware like in-memory processing~\cite{10.1109/ISCA45697.2020.00070,10.1145/3352460.3358284,9407100,10.1145/3466752.3480080}, MERCI~\cite{10.1145/3445814.3446717} takes an algorithmic way to memoize sub-query grouped results to reduce %the recommender system inference's 
memory pressure on the commodity server platform.

%\subsection{\arch DLRM Design}
%\niparagraph{\arch DLRM design.}
Different from \arch KV and \arch TX, we design \arch DLRM as an example of \textit{CPU-accelerator collaboration} for request processing.
Upon receiving a request from a client, the cc-accelerator first goes through the RPC stack, and then pass the request to the CPU through the ring buffer, where the request is parsed and transformed to model-ready input. Now, the input (request) is passed again to the cc-accelerator's APU, where the full inference, especially embedding reduction, is done. Finally, the cc-accelerator sends result (response) back to the client through the RNIC. 
Empirically, we observe that one CPU core with ~60\% usage can already keep up with the network and the cc-accelerator processing rate. 
In DLRM, not all memory accesses in a single query need to be serialized. 
Hence, in the APU, we issue 64 memory requests for each query's iteration so that the memory bandwidth can be fully utilized and the memory access latency can be hidden.  Lastly, the ALU is enhanced to support various aggregation operators (\eg, max/min/inner product).

\section{\arch Implementation and Evaluation Setup}
\label{sec:prototype}
We prototype \arch with a commercial system consisting of two Intel Xeon Gold 6138P CPUs at 2.0GHz~\cite{6138p} and 192GB DDR4 memory.
The configurations of the system are listed in \tabref{tab:res}.
Specifically, we use the in-package FPGA (Intel Arria 10GX@400MHz~\cite{arria10gx}) of the CPU to implement the \texttt{cpoll} mechanism and a cc-accelerator. 
The FPGA has a 64KB local cache and is connected to the CPU through a UPI link, typically used in NUMA systems to connect CPUs. 
The UPI link has one read channel and one write channel, each with 10.4GT/s bandwidth. It can also issue \texttt{sfence} signal and \texttt{cpoll} message.
For intra-machine communication, since HyperPlane's \texttt{QWAIT} and x86's user space \texttt{MWAIT} is unavailable on our platform, we use spin-polling for CPU to fetch requests in the request ring buffers from the cc-accelerators.

\begin{table}[!tb]
    \centering
    \caption{\arch testbed configurations.}
  \scriptsize
    \label{tab:res}
    \begin{tabular}{lr}
      \toprule
     \bf Intel Xeon 6138P CPU@2.0GHz &   \bf  \\
      \midrule
      \multicolumn{2}{l}{20 Skylake cores, hyperthreading enabled, running Ubuntu 18} \\ 
      \multicolumn{2}{l}{27.5MB shared LLC}\\
      \multicolumn{2}{l}{Six DDR4-2666 channels, 192GB DRAM in total}\\
      \midrule
     \bf In-package Intel Arria 10GX FPGA@400MHz &   \bf  \\
      \midrule
     \multicolumn{2}{l}{One UPI link to the CPU, 10.4GT/s (20.8GB/s)}\\
     %\multicolumn{2}{l}{CCIP over two PCIe 3.0x8 links to the CPU, 16GB/s}\\
     \multicolumn{2}{l}{64KB local cache}\\
     \multicolumn{2}{l}{ Resource usage: 11K (26\%) LUT; 130K (8\%) registers; 387 (14\%) BRAM blocks}\\
    %LUT usage& 11K (26\%) \\ 
    %Registers usage& 130K (8\%)\\
    %BRAM blocks usage& 387 (14\%) \\
      \midrule
     \multicolumn{2}{l}{\bf NVIDIA BlueField-2 DPU (Smart NIC) SoC}\\
      \midrule
     \multicolumn{2}{l}{2x25Gbps Ethernet ports, backed by ConnectX-6 Dx controller}\\
     \multicolumn{2}{l}{RDMA over converged Ethernet V2 (RoCEv2)}\\
     \multicolumn{2}{l}{Eight ARM A72 cores@2.5GHz, running Ubuntu 20}\\
     \multicolumn{2}{l}{6MB shared LLC}\\
     \multicolumn{2}{l}{16GB on-board DDR4-1600 DRAM}\\
     \multicolumn{2}{l}{One-sided RDMA for ARM to access the host memory}\\
      \bottomrule
    \end{tabular}
\end{table}

%The FPGA has a 64KB local cache (fixed by the Intel firmware) and is connected to the CPU via two types of coherent links. The first type is coherence-native UPI link, typically used in NUMA system to connect CPUs. It has one read channel and one write channel, each with 10.4GT/s bandwidth. It can also issue \texttt{mfence} signal and \texttt{cpoll} message. 
%The second type is two PCIe 3.0x8 links wrapped by CCIP protocol, which is also in the coherence domain. 
%Prior work~\cite{10.1145/3373376.3378482,10.1145/3445814.3446696} has shown that UPI's latency is shorter than PCIe's. 
%To maximize the throughput and avoid latency penalty, we designate the UPI link to the perform the most critical memory requests, such as \texttt{cpoll}-related data and data from the RDMA SQ handler to controller the RNIC. 
%Only when the UPI link is saturated will we use the PCIe links. 
%To hide the long PCIe link latency, we use them in a ``prefetching'' manner. That is, when observing query queueing in the application processing unit, we will issue memory requests for them via PCIe at first. Then, when it is the queued queries' turn to be processed, it can get the data immediately. 
We implement a round-robin algorithm in the scheduler.
The APU can support 256 outstanding requests. 
Each request buffer has 1024 entries. %, and a 4-byte counter is attached for \texttt{cpoll} region.
 We adopt the HERD's RPC protocol~\cite{196243,10.1145/2619239.2626299} for its simplicity, but any advanced RPC stack can also be applied~\cite{10.1145/3445814.3446696}.
The resource utilization numbers in \tabref{tab:res} reflect our \arch key-value store accelerator (\secref{sec:kvs}). The utilization results for the other applications we have built are similar, because $\sim80\%$ of used resource is for the coherence controller and the local cache, which are common components. 

The current implementation has two major limitations.
The first one is the performance of the coherence controller. As a soft design in the programmable fabric of the FPGA, it suffers from synthesis constraints and can perform at at most 400~MHz, incurring limited data access performance, which has also been observed by prior work~\cite{10.1145/3445814.3446696}). However, its counterparts on a regular server CPU can operate at $\sim$2~GHz~\cite{intelprivate}. We expect such infrastructural parts can be fixed by hard IPs in the future FPGA, offering performance of the accelerator's coherence controller comparable to that of the CPU's~\cite{stratix10dx}.
The second is that the FPGA lacks local memory that exists in the same coherence domain and has comparable capacity to CPU-attached memory. Consequently, most memory requests for application request processing will need to go through the cc-interconnect (due to the large working set of the target applications), similar to cross-NUMA memory access. 
Besides, the \texttt{cpoll} region must be pinned in the cc-accelerator's local cache.
We expect that such limitations will be effortlessly tackled when \arch is implemented in CXL-based devices~\cite{samsungcxl} or Enzian~\cite{enzian}. % as well in the near future -- in that case, the cc-accelerator can achieve higher throughput and lower latency.

Since the in-package FPGA is not extendable, to explore the potential of \arch's performance on future platforms, we also use a stand-alone Xilinx U280 FPGA card~\cite{u280} with 32~GB DDR4 memory and 8~GB HBM2 to emulate a cc-accelerator with local coherent memory~\cite{enzian,samsungcxl}.
Prior work~\cite{9114755} has shown that these two types of memory can achieve $\sim$36~GB/s and $\sim$425~GB/s throughput, respectively.
Specifically, we adapt the APU to the U280 card with either DDR4 or HBM2 controller. The application data is mapped and initialized in the FPGA's local memory. 
Rather than interacting with a real RNIC, we emulate arrival of RDMA requests by generating requests within the FPGA with the RDMA write rate measured on the testbed. 
We believe this emulation methodology offers correct and convincing results since coherence (of the application data) makes no big difference here after the data has been allocated and initialized in the FPGA-attached memory; during request processing, most memory traffic does not need to go across the cc-interconnect. 
For throughput experiments, we measure requests processed on the FPGA per second.
For latency experiments, we compute the emulated end-to-end latency by combining application processing time measured on the U280 with the average latency of the rest of the stack. Specifically, we measure the average latency from a request's generation to its completion on the U280, then add the average full-system end-to-end latency without an APU -- measured on the client machine. Note that this approach emulates average latency, so does not apply to tail latency measurements.
In the following sections, we notate the U280 DDR4-based results as ``\arch-LD (local DDR4)'' and HBM2-based results as ``\arch-LH (local HBM2)''.

Lastly, we use the NVIDIA ConnectX-6-based BlueField-2 DPU~\cite{bluefield2}
as the RNIC. It also provides eight ARM A72 cores at 2.5~GHz, which we use to compare against a conventional Smart NIC approach.

\section{Evaluation}
\label{sec:evaluation}

\subsection{Notification Latency: cpoll vs. Conventional Polling}
%After implementing 
To demonstrate the advantage of \texttt{cpoll} over conventional spin-polling,
%the \texttt{cpoll} mechanism described in Sec.~\ref{sec:cpoll}, 
we perform a local ping-pong test. 
Specifically, we allocate a 1~KB request buffer shared between the CPU and the FPGA.
The CPU first starts a timer to measure round-trip latency, writes the buffer's first byte, and then spins to poll the buffer's last byte. After detecting a change in the last byte,  the CPU stops the timer.
The FPGA polls or \texttt{cpoll}s the buffer's first byte. When it detects any change in the first byte, it immediately writes the buffer's last byte.  
With this setup, we test \texttt{cpoll} and polling with different polling intervals in FPGA cycles for 60K times. This is to measure notification latencies perceived by (1) the FPGA when the CPU sends a request with both \texttt{cpoll} and polling and (2) the CPU when the FPGA sends a request with polling. 
%For this test, first, both the CPU and the FPGA do polling every 15 FPGA clock cycles without any other traffic. As such, we can assume that the CPU-to-FPGA and FPGA-to-CPU portions a 50\%/50\% in the measured round-trip latency. The \textit{base value} here is $Avg(latency)/2$. 
%During the experiment, we also randomly issue memory requests every five cycles from the FPGA as the background noise to reflect a real scenario when processing the requests of memory-intensive applications.  
\begin{figure}[!b]
    \centering 
    \includegraphics[width=\linewidth]{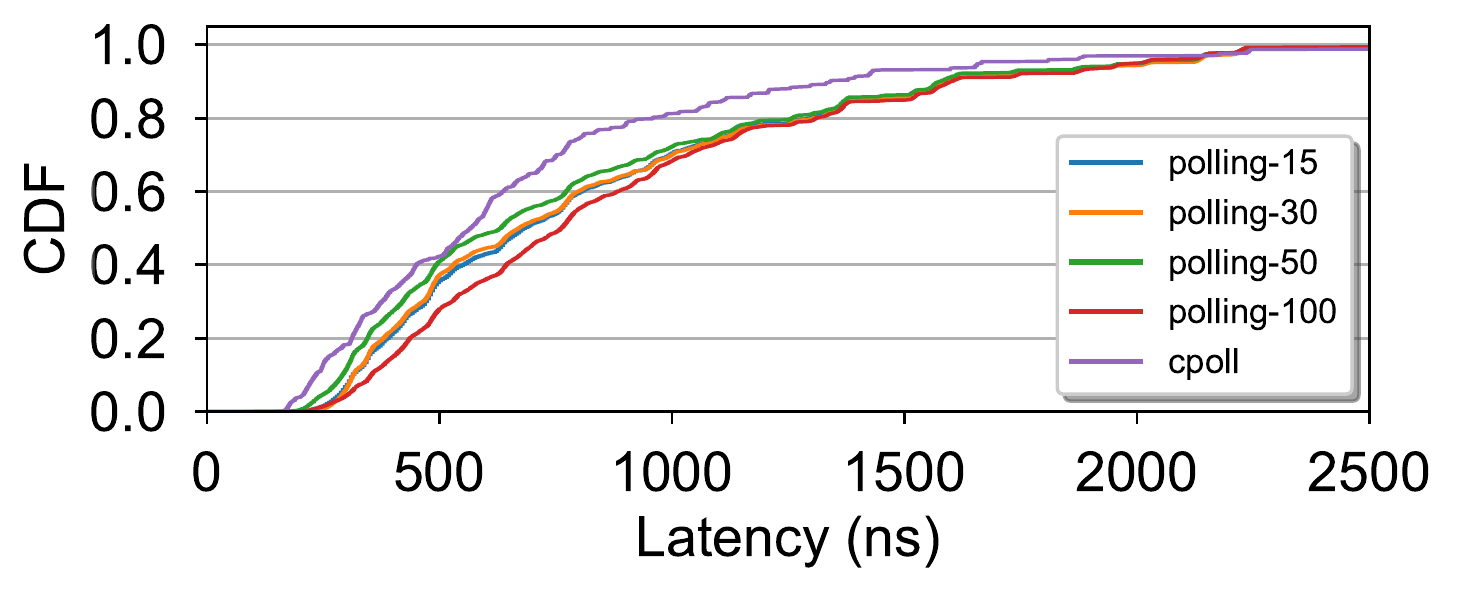}
    \caption{Latency distribution of \texttt{cpoll} and conventional polling with different polling intervals (cycles).}
        \label{fig:cpoll-exp}
\end{figure}

We plot the one-direction CPU-to-FPGA latency CDF in \figref{fig:cpoll-exp}. % (measured round-trip latency minus the constant \textit{base value}). 
First, \texttt{cpoll} always has a better average and tail latency than conventional polling, which can be as high as $\sim 30\%$.
Second, \texttt{cpoll} consumes less bandwidth of the interconnect (UPI in our case). Take polling-15 (15 FPGA clock cycles per polling) as an example: it may generate $64B * 400MHz/15\approx 1.6$GB/s traffic on the UPI link and coherence controller for a single request buffer, which affects the normal operations of the applications. 
Note that due to the frequency limitations of the FPGA, the latency's absolute value is not extremely low. 
Since the UPI link may only consumes $\sim$50ns latency~\cite{10.1145/3373376.3378509,intelprivate}, we expect that the coherence controller in the FPGA can be a hard-IP in future products to achieve higher performance~\cite{stratix10dx}.

\subsection{In-Memory Key-Value Store}
\label{sec:kv:exp}
Due to the limited availability of devices, we run \arch KV on one server and one client;\footnote{Since we use multiple cores (threads) on the client and the network bandwidth has been saturated, adding more clients will not affect the results.} see \secref{sec:disc} for scalability discussion.
We compare \arch with two state-of-the-art baselines: open-source two-sided RDMA-RPC (MICA-backed)~\cite{196243,10.1145/2619239.2626299} (noted as \textit{``CPU''}) and \textit{Smart NIC}~\cite{li2017kv,10.1145/3342195.3387519}. 
For \textit{CPU}, we use ten threads (cores) on the testbed to maximize the the KVS throughput.
Each thread is fed with requests by one client instance (each with two dedicated Skylake cores) on the client machine (also equipped with the BlueField-2 DPU). 
For \arch, we also use 10 client instances to feed requests that are processed on the same \arch accelerator. 
For \textit{Smart NIC}, we use DPU's \textit{all} eight ARM cores to emulate the behavior of the specialized hardware in KV-Direct~\cite{li2017kv} and StRoM~\cite{10.1145/3342195.3387519}.
The ARM cores process the request, which is sent from the client Intel CPU by RDMA. 
Besides, the ARM cores communicate with the server host through RDMA (direct verbs) for necessary data retrieval.
Admittedly, the ARM core is not as efficient as the specialized FPGA designs~\cite{li2017kv,10.1145/3342195.3387519} when processing KVS and accessing host memory.
However, the ARM cores' frequency is $\sim10\times$ higher, alleviating the efficiency gap. 
Also, we use direct verbs~\cite{mlx5dv} to minimize the overhead imposed by the RDMA software stack.
Based on our measurement, when running KVS entirely on the Smart NIC's on-board DRAM, the eight ARM cores' peak throughput is equivalent to six Intel CPU cores.

We pre-load 100M key-value pairs (64~B size, $\sim$7~GB memory in total) and then access them using uniform and Zipfian 0.9 distributions. 
We test two types of workloads: read-intensive (100\% GET), and write-intensive (50\% GET, 50\% PUT). 
Note that the MICA-based mechanism~\cite{196243,10.1145/2619239.2626299}, which we use in this experiment, eliminates the concurrency issue (\ie, only allowing the “owner core” to read/write the data partition). As such, the performance of heavy PUT workload does not differ much from the GET only workload, which is aligned with the results in prior work~\cite{li2017kv,196243}. 

In \textit{CPU} and \arch, both the hash table and key-value pairs are stored in the host memory; in \textit{Smart NIC}, we allocate a 512~MB space on the DPU's on-board DRAM as the cache to store the most recently accessed hash entries and key-value pairs. The cache-total ratio (512~MB~:~7~GB) is roughly the memory capacity ratio (16~GB~:~192~GB).
We also test the impact of batching. 
In \textit{CPU} and \textit{Smart NIC}, batching means processing requests in a batch to improve the memory access efficiency~\cite{179747}. 
In \arch, since the APU can already exploit the memory-level parallelism across requests~\cite{9407097,kocberber2013meet,10.1145/3373376.3378497,9407194}, there is no need for request batching. Hence, we batch the doorbell signals to the RNIC~\cite{196243} when posting RDMA operations for response.
These settings and configurations resemble prior work~\cite{10.1145/3445814.3446696,196243,li2017kv,179747}.

\begin{figure}[!t]
    \centering 
    \includegraphics[width=\linewidth]{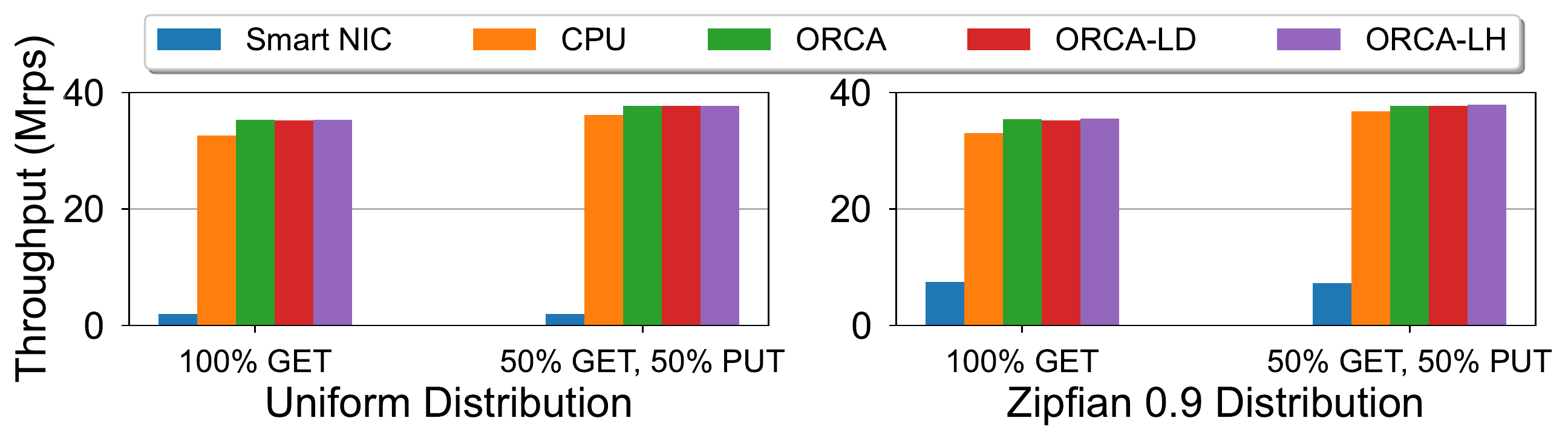}
    \caption{Peak throughput performance of different KVS designs. The batch size of 32 is applied.}
    \label{fig:mica-tp}
\end{figure}

\begin{figure}[!b]
    \centering 
    \includegraphics[width=\linewidth]{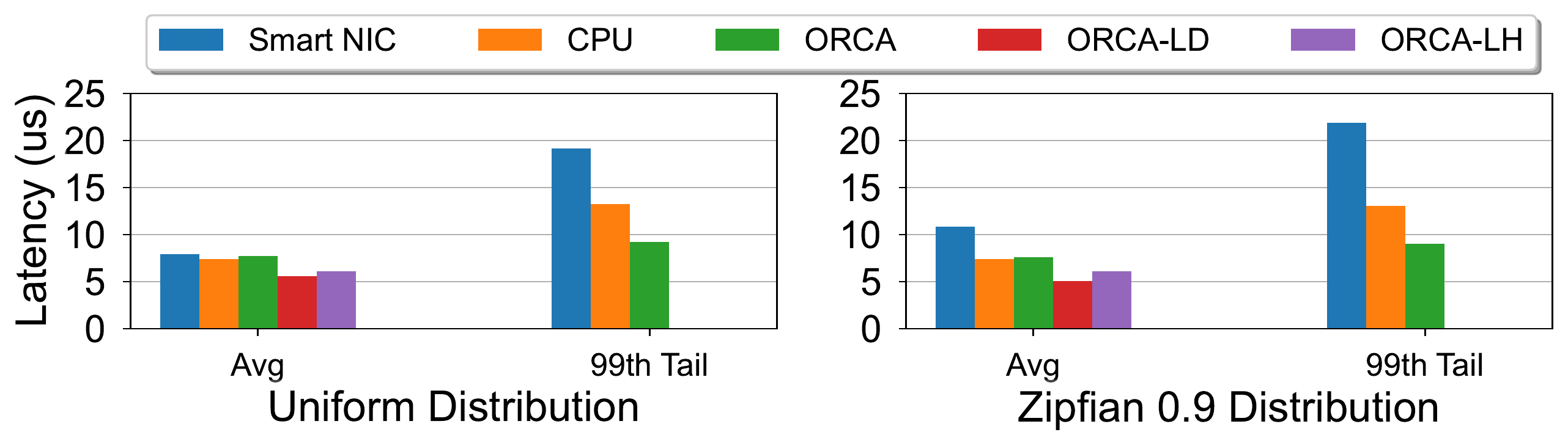}
    \caption{Latency performance of different KVS designs on the 100\% GET workload. The batch size of 32 is applied. \arch-LD/LH's tail latency is inapplicable.}
    \label{fig:mica-lat}
\end{figure}

We first show each design's peak performance (with batch size 32). 
Regrading the throughput in \figref{fig:mica-tp}, we first find that the request distribution significantly affects \textit{Smart NIC}'s performance. 
\textit{Smart NIC}'s throughput with uniform distribution (\ie, more than 90\% memory accesses are to the host via PCIe) is 27.2\%--28.6\% of that with a skewed Zipfian distribution (\ie, most memory accesses are local). And even the throughput with Zipfian distribution is only $\sim$60\% of that with pure on-board memory accesses. 
On the other hand, the distribution does not affect \textit{CPU} and \arch's performance, since even with the Zipfian distribution, the KVS's memory footprint is still larger than the CPU or FPGA's cache. 
Second, we observe that \arch's peak throughput is 2.3\%$\sim$8.3\% higher than \textit{CPU}. 
This is because the peak KVS throughput is bounded by the network bandwidth now, and \arch's one-sided RDMA performs a little better than \textit{CPU}'s two-sided RDMA, which is aligned with prior studies~\cite{225980,10.1145/3319647.3325827}. 
The throughput of \arch-LD and \arch-LH can prove this as well -- extra memory bandwidth does not help improve the performance (in fact, the UPI link is not saturated), since the network has reached its limit.

Regarding the latency, taking the 100\% GET workload as an example (\figref{fig:mica-lat}), the \textit{Smart NIC}'s performance is again affected by the request distribution, since the PCIe link adds significant latency, even if the accesses are batched.
Meanwhile, we observe that \arch's average latency is a bit higher than \textit{CPU}. This is mainly because, unlike \textit{CPU}, \arch needs to access data through the UPI link, adding more time on the request processing critical path. 
This deficiency is overcome with \arch-LD/LH's accelerator-attached memory -- it only goes through the UPI to interact with the RNIC. 
Note that due to HBM's nature, \arch-LH has a higher average latency than \arch-LD since the workload is not bounded by memory bandwidth now. 
For tail latency, \arch is 52.0\% lower than \textit{Smart NIC} and 30.1\% lower than \textit{CPU}, because it not only significantly remove the PCIe overhead, but also has more stable behavior than the CPU core, whose performance is affected by multiple factors like OS scheduling and CPU resource contention. 

\begin{figure}[!t]
    \centering 
    \includegraphics[width=\linewidth]{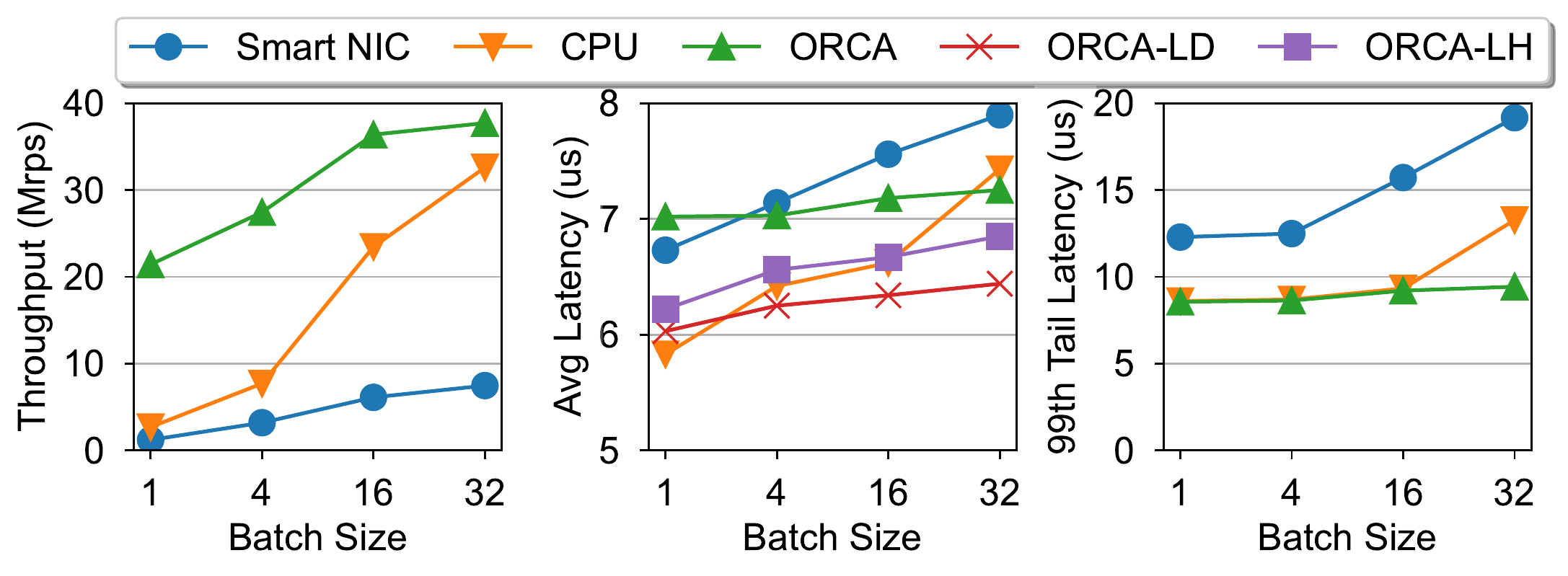}
    \caption{The impact of batch size on throughput and latency (100\% GET workload, Zipfian 0.9 distribution). \arch-LD/LH's tail latency is inapplicable.}
    \label{fig:mica-batch}
\end{figure}

We also investigate the impact of the batch size on each design and demonstrate the results in \figref{fig:mica-batch} (since the KVS throughput is now network-bound, we do not include \arch-LD/LH's throughput as they are the same as \arch).  
For \textit{CPU} and \textit{Smart NIC}, batching can significantly improve their throughput performance (\ie, $\sim12\times$) -- by batching the data accesses across requests, the memory bandwidth is efficiently utilized and the memory/PCIe latency is hidden. 
\arch's throughput also benefits from batching (\ie, $\sim2\times$), which is because of the reduction of MMIO-based doorbell access~\cite{179050,10.1145/3352460.3358278,196243}, and MMIO's surrounding \texttt{sfence} signals from the \arch cc-accelerator, which is relatively expensive.  
On the other hand, unlike \textit{CPU} and \textit{Smart NIC}, \arch's latency sub-linearly increase with the batch size. 
This is because \arch does not need to wait for the batch size of arrived requests to start processing, and the RNIC may execute the WQE promptly before the doorbell is rang~\cite{mlxprm}.

\begin{table}[!t]
    \centering
    \caption{Overall power efficiency of different KVS approaches with GET operations in uniform distribution.}
    \label{tab:power}
    \begin{tabular}{cccc}
        \toprule
          & \textit{CPU} & \textit{Smart NIC} & \arch \\
          \midrule
    Kop/W & 130.4    &   25.2        &     188.7\\
    \bottomrule
    \end{tabular}
\end{table}

Finally, we use Intel RAPL interface~\cite{rapl} (for CPU and DIMMs), IPMI tool (for the entire server box), and the FPGA's firmware (for the FPGA chip) to measure the power consumption of the evaluated approaches. Take the case in \figref{fig:mica-tp} as an example.
We find that the \textit{CPU} and \textit{Smart NIC}'s Intel/ARM CPUs consume $\sim$90 Watts and $\sim$15 Watts respectively when fully loaded, while \arch's FPGA power is in the range of 24--27W to achieve the peak throughput. This demonstrates \arch cc-accelerator's $\sim3\times$ power efficiency than the beefy Intel CPU to achieve comparable performance, leading to $\sim38\%$ power consumption reduction of the entire server box, as demonstrated in \tabref{tab:power}.

\subsection{Distributed Transaction with NVM-based Chain Replication}
\label{sec:trans:exp}

According to the HyperLoop paper~\cite{10.1145/3230543.3230572}, HyperLoop mechanism always outperforms CPU-based chain replication implementations, especially in the multi-tenant cloud environment, so we compare \arch only with HyperLoop.
Same as the HyperLoop paper, we adopt RocksDB~\cite{rocksdb}, a persistent key-value database, to use the NVM as the persistent storage medium and to apply \arch and HyperLoop.
Since HyperLoop modifies RNIC's firmware, which is not open-source, we use the ARM cores on the DPU to emulate its behavior.  
Since the (Skylake) CPU with in-package FPGA does not support Intel Optane DIMM, we emulate NVM's behavior by adding latency and throttling memory bandwidth in the FPGA and the ARM emulation program.
We follow the NVM's characteristics in recent Optane-based studies~\cite{246192,10.1145/3419111.3421294} to calibrate our emulation. 
We disable DDIO on the server.

Having the same device limitation, we run the experiments on one client and one server. 
We make use of the two ports on the DPU to have two replica machines (instances) in the same physical server, and the transaction will be forwarded across the two ports, as shown in \figref{fig:chain}. 
The client's host CPU will initiate a transaction and send it to the server's port 0 (\circled{1}). 
The corresponding processing unit instance (either a DPU ARM core or \arch) will forward the transaction to the client's DPU ARM via port 0, which is attached to a RocksDB instance (\circled{2}). 
The client's DPU ARM simply routes the transaction to the server's port 1, where another RocksDB instance (and the corresponding processing unit) serves as the second replica machine (\circled{3}). 
Finally, the transaction will be sent back to the client's host CPU (\circled{4}). 
According to our measurement, the ARM-based routing will add 2$\sim$3$\mu s$ overhead, which resembles the network latency in the real datacenter. 

We initiate the RocksDB instance with 100K key-value pairs and issue 100K transactions from the client to measure the end-to-end latency. 
We test two key-value pair sizes (64B and 1024B) and two types of transactions with different \textit{(read, write)} counts (\textit{(0,1) and (4,2)}, representative in real-world transactional systems~\cite{199315}).
Since the \arch Tx and HyperLoop has the same mechanism for pure-read transactions, we exclude such transactions from the evaluation.

\begin{figure}[!t]
    \centering 

    \includegraphics[width=\linewidth]{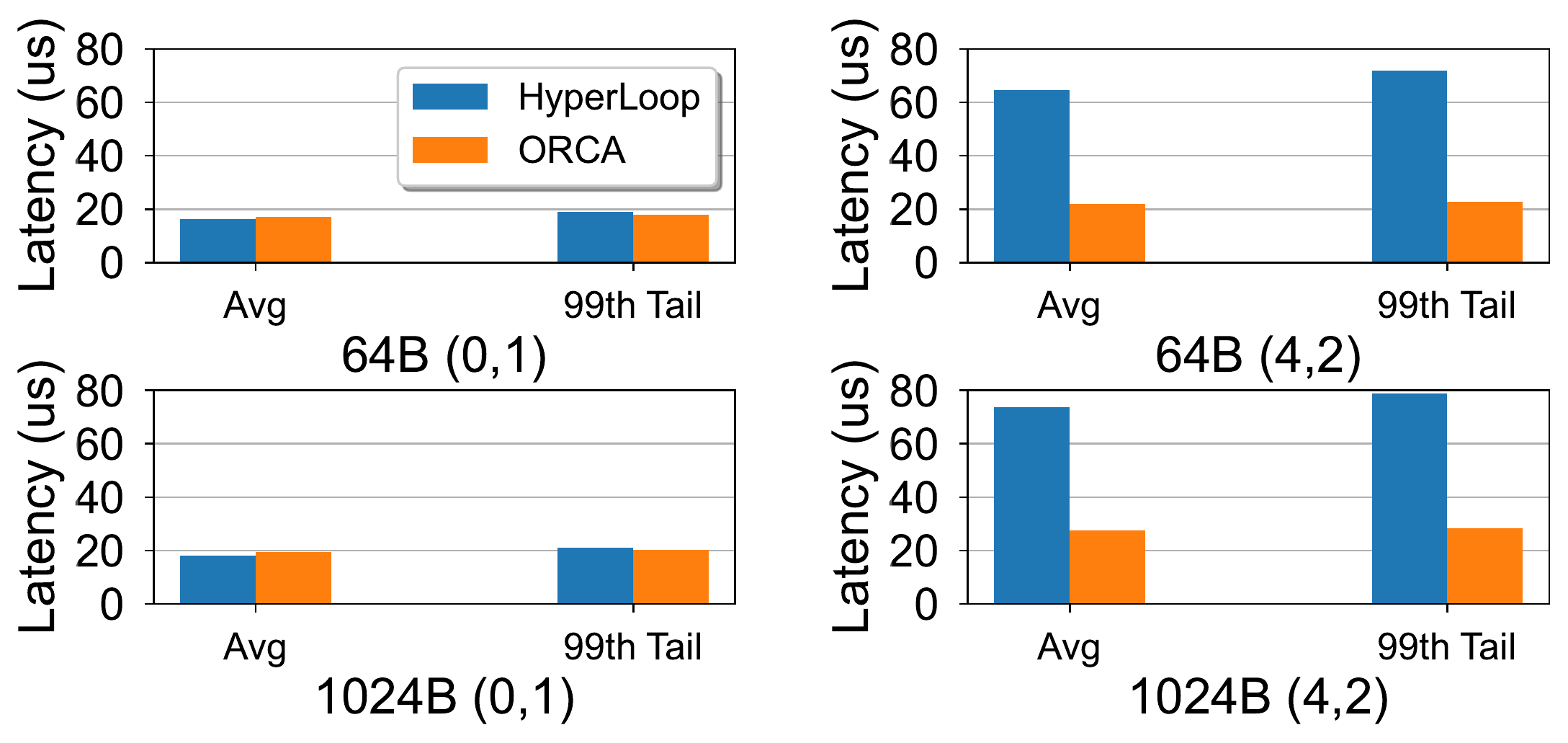}
    \caption{Latency comparison with different key-value pair size and  transactions with different numbers of \textit{(read,write)}.}
    \label{fig:hl-lat}
\end{figure}

\begin{figure}[!b]
    \centering 
    \includegraphics[width=\linewidth]{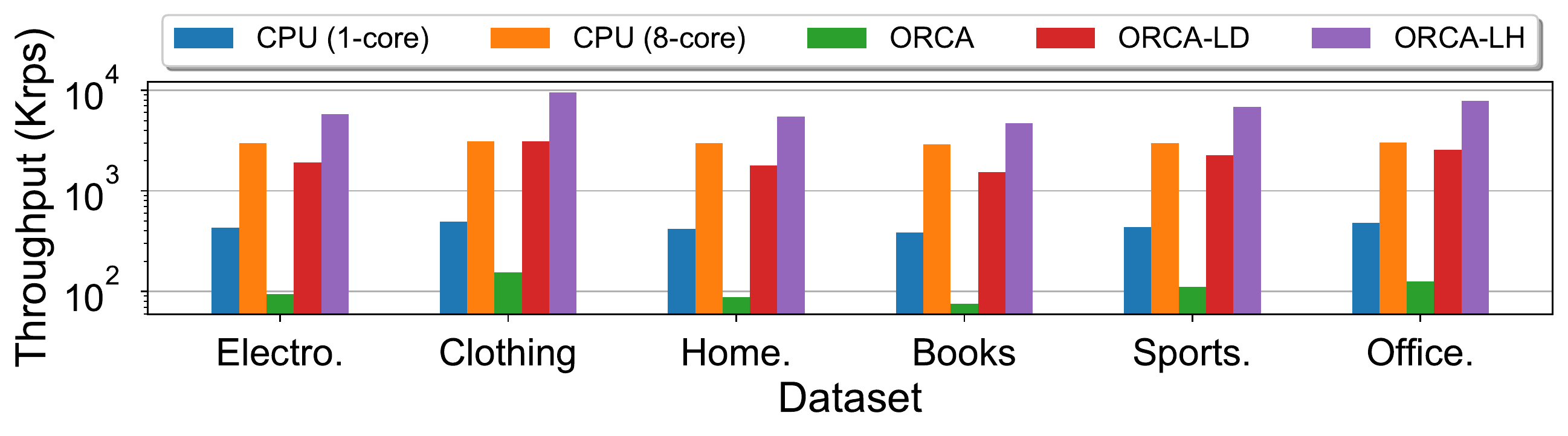}
    \caption{MERCI-based DLRM inference throughput on the Amazon Review dataset.}
    \label{fig:merci}
\end{figure}

We demonstrate the latency results of HyperLoop and \arch in \figref{fig:hl-lat} (note that since the transactions are issued by the client one by one, the latency improvement can also reflect the throughput improvement).  
For the \textit{(0,1)} transaction, \arch's performance does not differ from HyperLoop, since they experience the same overhead -- one PCIe round-trip per replica machine and one round-trip over the 2-machine replication chain, and \arch may even be a bit (less than 3\%) slower than HyperLoop since it also has the overhead of UPI link. 
However, when the transaction contains multiple operations, \arch begins to show its advantage. 
Unlike HyperLoop, \arch's client only needs to issue one combined transaction request to the replication chain, and the \arch accelerator will handle the transaction execution and chain replication protocol itself in a near-data manner -- still one PCIe round-trip per machine and one round-trip over the chain.
This saves the network and PCIe latency, offering a 63.2\%$\sim$66.8\%  reduction for average end-to-end latency and 64.5\%$\sim$69.1\% for 99th tail latency. 
Note that \arch's latency can be further reduced with accelerator (FPGA) that can directly attach NVM~\cite{stratix10dx}.

\subsection{DLRM Inference}
\label{sec:recom:exp}

We follow MERCI~\cite{10.1145/3445814.3446717} and facebook's DLRM model~\cite{naumov2019deep} to build our test program. Since their open-source implementations only include single-machine version, we extend them to a RDMA-based end-to-end datacenter application (the networking part is similar to the optimized HERD~\cite{196243}) to reflect the real datacenter environment. 
The client send the query request to the server for inference. 

We follow the configurations in the MERCI paper~\cite{10.1145/3445814.3446717} to perform the evaluation.  
We compare \arch's performance with the CPU-based software version on the popular Amazon Review dataset~\cite{10.1145/2872427.2883037} (electronics, clothing-shoe-jewelry, home-kitchen, books, sports-outdoors, and office-products).
Both the native embedding reduction and MERCI reduction are evaluated. 
The data is clustered and loaded into embedding tables and memoization tables by the CPU in MERCI's way. 
The embedding dimension is set to the default value of 64; for MERCI reduction, we build memoization tables with 0.25$\times$ the size of the original embedding tables.

Since each query's length (number of features) in a dataset is diverse, it is unfair to measure the per-query latency, so we only measure the throughput. 
For brevity, we only show the results of inference with MERCI reduction in \figref{fig:merci} because the ones with native reduction show the same trend.
For CPU-based version, MERCI scales linearly until eight cores (threads), which is bounded by the host memory bandwidth. 
For \arch, however, we find poor throughput performance over all the datasets -- only 19.7\%$\sim$31.3\% throughput of a single CPU core.
After further analysis, we find this phenomenon is because
(1) unlike KVS, the nature of the embedding reduction in DLRM (\ie, pure and dense memory accesses within nested ``for'' loops, few branches, thousands of memory accesses per query) makes it relatively efficient on the CPU core -- the instruction window and the load-store queue can be fully utilized;
(2) the CPU core can leverage the entire bandwidth of the memory channels (\ie, $\sim$120GB/s on our testbed) with good parallelism, while the \arch cc-accelerator can only leverage the cc-interconnect's bandwidth, and the memory requests have to be issued serially from the FPGA's wimpy coherence controller;
(3) the compute-intensive fully connected layer is relatively lightweight in the model, making \arch's hardware acceleration only a small portion in the end-to-end inference.
Hence, with higher frequency and memory bandwidth, CPU outperforms the \arch cc-accelerator.
This deficiency can be solved by \arch-LD and \arch-LH. 
Fully utilizing the two DDR4 channels ($\sim\frac{1}{3}$ of the CPU memory channels' bandwidth), \arch-LD is able to achieve 52.8\%$\sim$95.3\% throughput of the eight CPU cores. 
Furthermore, the 32-channel HBM2 eliminates the memory bandwidth bottleneck in the reduction, leading to \arch-LH's $1.6\times\sim3.1\times$ throughput improvement over the CPU, and the RDMA network becomes the limiting factor for higher end-to-end throughput. 
Note that, CPU with integrated HBM2~\cite{intel-hbm} in the near future may also achieve similar throughput compared to \arch-LH; but similar to KVS (see \secref{sec:kv:exp}), even with the same memory bandwidth (and thus the inference throughput), \arch cc-accelerator shows better power efficiency over the CPU-based reduction.

%\input{kvstore}
%\input{trans}
%\input{recom}
% !TEX root = paper.tex
\section{Discussion}
\label{sec:disc}
%Due to the limited availability of devices including RDMA NICs and integrated FPGA, 
%We are constrained to implement and evaluate \arch on a small group of machines with 25Gbps network. 
%However, we argue that this will not affect \arch's scalability within higher network bandwidth and larger clusters. 
%Also, \arch can be extended to support multiple applications simultaneously.

%\subsection{Scalability with Faster Network}
\niparagraph{Scalability with faster network.}
As the speed of network is growing fast, a critical question is whether \arch will keep up with the speed of future network (\eg, 400Gbps). 
First, as mentioned before, the UPI's bandwidth is not saturated in \arch KV and \arch TX. 
Furthermore, as demonstrated by \arch DLRM, accelerator-attached memory with comparable capacity to the CPU~\cite{enzian,samsungcxl,stratix10dx} will further librate the  bandwidth of the cc-interconnect from application-related memory requests. Hence, the cc-interconnect can better serve the  RNIC/CPU-cc-accelerator interaction. 
This means \arch will be bottlenecked by the network bandwidth and can achieve higher performance with newer network technologies (also note that the cc-interconnect performance will evolve as well). 

%\subsection{Scalability with Larger Cluster}
\niparagraph{Scalability with larger cluster.}
%\label{sec:scale-out}
Prior work has stated that one-sided RDMA operations, relying on reliable connection, cannot scale well to %hundred-level 
hundreds of machines~\cite{225980,199315,196243}. 
This is mainly due to the limited 
size of on-chip cache of the RNIC, which stores the necessary connection information. 
However, this has been alleviated with the advance %of 
in technologies~\cite{10.1145/3319647.3325827}. 
For instance, as the latest NIC product, the blueField-2 DPU has $\sim$10MB cache for its 8-core ARM system on the SoC~\cite{bluefield2}, which also integrates the ConnectX-6 controller. 
We expect that the ConnectX-6 part on the SoC has comparable cache size (\eg, 5$\sim$10MB) for connections. %(NVIDIA does not disclose the exact cache size). 
Based on the previous calculation~\cite{225980}, such on-chip cache can support %>10K 
more than 10K connections without performance loss by cache miss, which is enough for most clusters in modern datacenters. 
Also note that, \arch design itself does not worsen the RDMA's scalability problem.
%Meanwhile, threads/processes on the same end-host can share the request/response buffer (and the underlying QPs) with pure software approach with little/no performance loss~\cite{10.1145/3477132.3483576,10.1145/3302424.3303968,10.1145/3132747.3132762}, which further mitigates the scalability concerns of the RDMA connections and buffers.

%Another potential limiting factor %of 
%for scale-out is the size of the \texttt{cpoll} region, which has been discussed in \secref{sec:cpoll}. 

%\subsection{Concurrent Execution of Multiple Applications}

\iffalse
\niparagraph{Concurrent execution of multiple applications.}
In cloud environments, it is desirable to run multiple applications on the same physical platform, including the accelerator. 
Multiple works have explored the virtualization of accelerators including FPGA-based ones in cloud environments 
~\cite{222551,zha2021isca,zha2020asplos,zha2018asplos}, especially in the coherence domain~\cite{10.1145/3373376.3378482}. 
They make it possible to split accelerators and host memory resource to multiple instances (and thus applications). 
%Second, accelerators like QEI~\cite{9407097} and Livia~\cite{10.1145/3373376.3378497} leverage the common execution patterns to accelerate various memory-centric applications/routines on unified hardware.
\arch can by extended by this line of research to for better multi-application support.
\fi
% !TEX root = paper.tex
\section{Related Work} 
\label{sec:related}
\niparagraph{Accelerating $\mu s$-scale datacenter applications with emerging devices.}
In modern datacenters, commodity networking devices, especially programmable switches and Smart NICs, have been leveraged to accelerate datacenter applications. 
Such approaches include caching~\cite{10.1145/3132747.3132764,liu2017incbricks,li2017kv,pmnet}, compute offloading~\cite{sapio2019scaling,sapio17:_in_networ_comput_dumb_idea,sharp,lerner2019case,10.1145/3230543.3230555,li2017kv,179402,10.1145/3422604.3425923,10.1145/3342195.3387519}, protocol offloading~\cite{li:_pegasus,10.1145/3132747.3132751,211261,10.1145/3387514.3405857,10.14778/3368289.3368301,ports15:_desig_distr_system_using_approx,li16:_fast_replic_nopax,dang16:_networ_hardw_accel_consen,dang18:_p4xos,264822,10.1145/3230543.3230572,10.1145/3477132.3483561,10.1145/3477132.3483555}, load balancing~\cite{227794,10.1145/2872362.2872367}, \etc.
However, due to the limited memory capacity of such devices (\eg, \textit{O(10MB)} of on-chip SRAM for programmable switches~\cite{intelprivate,10.1145/3387514.3405855} and \textit{O(10GB)} of on-board DRAM for Smart NICs~\cite{bluefield2,10.1145/3341302.3342079}), they may fall short of efficiently handling datacenter applications requiring large memory footprints.
To tackle such a challenge, TEA~\cite{10.1145/3387514.3405855} proposes to let the switch retrieve data from the affiliated servers by low-latency RDMA read.
TEA uses algorithm design (\ie, linear probe in its hash table) to reduce the required network round-trips to get the desired data, but it cannot always do so for all applications/data structures (\eg, B-tree based KVS~\cite{258947}).

%\niparagraph{Integrated NIC/accelerator for low intra-machine communication.}
To alleviate the interconnect (PCIe) overhead, a line of research integrates the entire NIC or accelerator to/near the CPU package~\cite{10.1145/3352460.3358278,10.1145/3297858.3304070,10.1145/2749469.2750415,10.1145/2541940.2541965,273715,10.1145/2485922.2485926,10.1145/3466752.3480051,10.1145/3466752.3480055,clio}. 
Enjoying the benefit of fast NIC-core interaction, this approach has (1) high cost of designing and manufacturing them and (2) low flexibility of usage and maintaining.  
For example, a typical NIC ASIC's die area can be more than 60$mm^2$~\cite{intelprivate}, which is roughly the area of four server CPU cores~\cite{skx-arch}.
Also, to upgrade the datacenter network infrastructure, the entire server CPU needs to be replaced, leading to high total cost of ownership (TCO). 
Dagger~\cite{10.1145/3445814.3446696} also leverages cc-accelerator for NIC design, but it still involves the server CPU for request processing, and only uses the cc-interconnect for its lower latency over PCIe.

Compared to these works, \arch takes a modularized design with low TCO, while still leveraging cache coherence feature to more efficiently handle applications with irregular/uniform memory access patterns.
Besides, for workloads with highly-skewed access patterns, \arch only adds one PCIe round-trip to the end-to-end latency than its counterparts.
They are complementary and thus they can work together.  

There is prior effort on accelerating applications in datacenter with cc-accelerator~\cite{10.1145/3445814.3446713,10.1145/3373376.3378482,10.1145/3445814.3446696,7966689,10.1145/3035918.3035954,10.1109/ISCA45697.2020.00083}, but they either accelerate single-machine applications/operations or a specific routine/layer in the system. \arch takes a holistic cross-stack approach to achieve end-to-end datacenter application offloading at $\mu s$-scale.

\niparagraph{NIC-host co-design framework.}
With the growing popularity of the Smart NIC, researchers have developed frameworks to schedule/offload datacenter applications to the Smart NIC's 
processors~\cite{234944,10.1145/3341302.3342079,222623}. 
However, they are still constrained to separate the Smart NIC and host's memory to two domains and suffer from a high cost of communications over PCIe links when memory-intensive code segments are offloaded.
\arch tackles these challenges with its unique near-data processing capability, while keeping the networking part offloaded on the RNIC for low cost and flexibility. 
\arch can also be included to these co-design frameworks as another compute resource.
Lynx~\cite{10.1145/3373376.3378528} proposes Smart NIC-based communication offloading for accelerator-rich systems, and FlexDriver~\cite{10.1145/3503222.3507776} proposes PCIe-based NIC control by accelerator. 
\arch takes one step further to let the client directly communicate with the accelerators, which also controls the NIC more efficiently in the coherence domain.
% !TEX root = paper.tex
\section{Conclusion}
\label{sec:concl}
We present \arch, a holistic system design to offload modern $\mu$s-scale datacenter applications.
It leverages the RDMA and cc-accelerator technologies to achieve high throughput and low latency performance with the CPU's involvement only when necessary. 
We apply \arch to three representative datacenter applications, each with its unique characteristics. 
Our evaluation on a real system shows that, compared to the CPU-based software solutions and the (Smart) NIC offloading solutions, \arch is able to achieve better and more stable performance with higher power efficiency.

\clearpage 

%%%%%%% -- PAPER CONTENT ENDS -- %%%%%%%% 

%%%%%%%%% -- BIB STYLE AND FILE -- %%%%%%%%
%\bibliographystyle{plain}
\bibliographystyle{IEEEtranS}
\bibliography{references}
%%%%%%%%%%%%%%%%%%%%%%%%%%%%%%%%%%%%

\end{document}